\documentclass[onecolumn]{aa} 

\usepackage[T1]{fontenc}
\usepackage[utf8]{inputenc}
 \usepackage{tgbonum}

\usepackage{booktabs}
\usepackage{graphicx}
\usepackage{amsmath}
\usepackage{amssymb}
\usepackage{txfonts}
\usepackage{lineno}

\defcitealias{protection1991icrp}{ICRP 60 (1991)}
\defcitealias{icru1980}{ICRU 33 (1980)}
\defcitealias{valentin2007icrp}{ICRP 103 (2007)}
\defcitealias{dietze2013icrp}{ICRP 123 (2013)}
\defcitealias{berger1984report}{ICRP 37 (1984)}

\newcommand*\mean[1]{\overline{#1}}

\begin{document}
{\fontfamily{qpl}\selectfont

\title{Revisiting the cosmic-ray induced Venusian radiation dose in the context of habitability}

  \author{Konstantin Herbst\inst{1}
          \and
          Sa\v{s}a Banjac\inst{1}
          \and
          Dimitra Atri\inst{2}
          \and
          Tom A. Nordheim\inst{3}
          }

   \institute{Institut f\"ur Experimentelle and Angewandte Physik, Christian-Albrechts-Universit\"at zu Kiel (CAU), 24118 Kiel, Germany\\
              \email{herbst@physik.uni-kiel.de, banjac@physik.uni-kiel.de}
         \and
            Center for Space Science, New York University Abu Dhabi, PO Box 129188, Abu Dhabi, UAE
        \and
             Jet Propulsion Laboratory,California Institute of Technology, Pasadena, CA 91109, USA}
\date{}
\authorrunning{Herbst et al.}
\titlerunning{Revisiting the Venusian radiation dose}
 
\abstract
{Cosmic rays (CRs), which constantly bombard planetary magnetic fields and atmospheres, are the primary driver of atmospheric spallation processes. The higher the energy of these particles, the deeper they penetrate the planetary atmosphere, and the more likely interactions become with the ambient atmospheric material and the evolution of secondary particle showers.}
{As recently discussed in the literature, CRs are the dominant driver of the Venusian atmospheric ionization and the induced radiation dose below $\sim$100 km. 
In this study, we model the atmospherically absorbed dose and the dose equivalent to the effect of cosmic rays in the context of Venusian habitability.
}
{The Atmospheric Radiation Interaction Simulator (AtRIS) was used to model the altitude-dependent Venusian absorbed dose and the Venusian dose equivalent. For the first time, we modeled the dose rates for different shape-, size-, and composition-mimicking detectors (phantoms): a CO$_2$-based phantom, a water-based microbial cell, and a phantom mimicking human tissue.}
{Based on our new model approach, we give a reliable estimate of the altitude-dependent Venusian radiation dose in water-based microorganisms here for the first time. These microorganisms are representative of known terrestrial life. We also present a detailed analysis of the influence of the strongest ground-level enhancements measured at the Earth's surface, and of the impact of two historic extreme solar events on the Venusian radiation dose. Our study shows that because a phantom based on Venusian air was used, and because furthermore, the quality factors of different radiation types were not taken into account, previous model efforts have underestimated the radiation hazard for any putative Venusian cloud-based life by up to a factor of five. However, because we furthermore show that even the strongest events would not have had a hazardous effect on putative microorganisms within the potentially habitable zone (51 km - 62 km), these differences may play only a minor role.} 
{}

\maketitle

\section{Introduction}
\label{sec:1}
As discussed in the literature, planetary atmospheres within our Solar System are exposed to a harsh radiation environment consisting of, for example, the solar wind, ultraviolet- and X-ray radiation, and cosmic rays (CRs) of Galactic or solar origin. The atmospheres may also include magnetospheric particle precipitation.

According to  \citet{Reames-1999}, solar energetic particles (SEPs) are accelerated in solar flares, coronal mass ejections, or at interplanetary shocks. Galactic cosmic rays (GCRs), however, are most likely produced by diffusive shock acceleration at the shocks of supernova remnants \citep[see, e.g.,][]{Hillas-2005, Buesching-etal-2005}. Thus, because of their different origins and acceleration mechanisms, the energy distribution of these two components differs significantly. While SEPs exceed the GCR flux by four orders of magnitude at lower energies (several keV), a steep decrease at energies above several MeV occurs \citep[see, e.g.,][]{Herbst-etal-2015}. Only in case of extreme SEP events that are detectable on the Earth's surface, so-called ground-level enhancements (GLEs), the spectrum exceeds energies of hundreds of MeV to a few GeV. In addition, differences in shape and composition of these GLE events have been reported \citep{Schmelz-etal-2012}. Although GCRs with energies of up to several TeV enter the heliosphere \mbox{omnidirectionally}, the heliospheric magnetic field (HMF), which is carried outward with the solar wind, modulates the GCR flux over a solar cycle.

When they are present, CRs encounter yet another magnetic filter in the form of the planetary magnetic field before they impinge on planetary atmospheres. There, CRs can be deflected by the Lorentz force, which leads further modulates the CR flux. However, CR interactions with atmospheric neutrals are the primary driver of the planetary low-altitude atmospheric ionization. While low-energy CRs lose most of their energy due to elastic collisions with atmospheric neutrals or ionization of the upper atmosphere, CRs with energies above $\sim$ 1 GeV can induce extensive secondary particle cascades by undergoing inelastic scattering with atmospheric nuclei before they are stopped and absorbed. In the case of the latter, the first generation of secondary particles is created, which mostly consists of mesons ($\pi^{\pm}$ and $\kappa^{\pm}$), nucleons, gamma particles, and nuclear fragments. Secondary particles with high energies may also be able to interact further, producing more secondaries that lead to the evolution of an atmospheric cascade consisting of the hadronic branch (neutrons and protons), the electromagnetic branch (electrons, positrons, and photons), and the muonic branch.
\begin{figure}[!t]
\centering
\includegraphics[width=0.7\columnwidth]{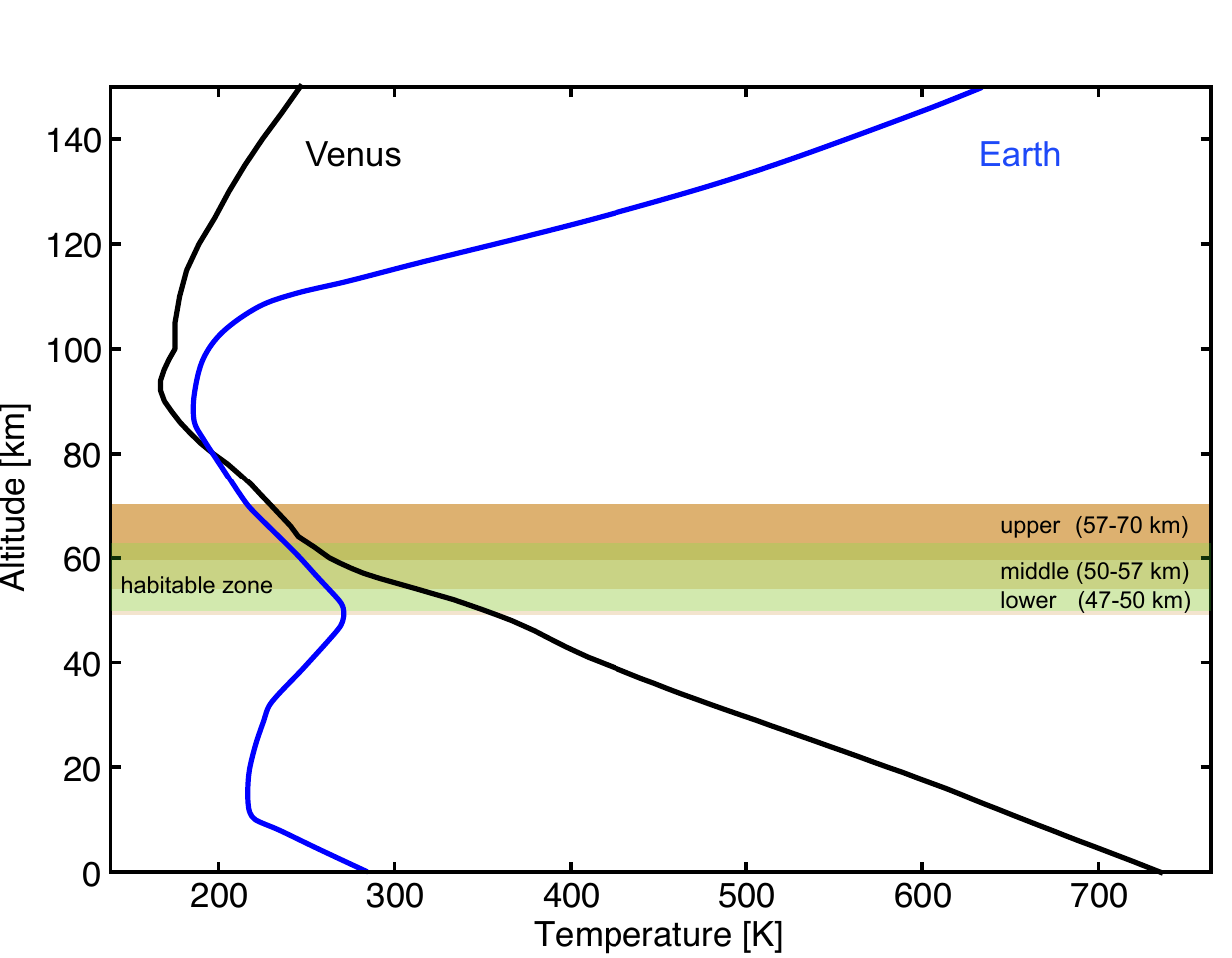}
\caption{Adopted atmospheric altitude-dependent temperature profile of Earth \citep[blue line, based on the NRLM-SISE00 model, see][]{picone2002nrlmsise} and Venus \citep[black line, based on the VIRA model, see][]{Keating-etal-1985, Seiff-etal-1985}.}
\label{fig:1}
\end{figure}
With increasing atmospheric depth, the secondary particle cascade evolves until the so-called Pfotzer maximum \citep{Pfotzer-1936} is reached, after which the average energy of secondary particles is not sufficient to further drive the production of new secondary particles. Consequently, the cascade starts to diminish after this point is reached. The Pfotzer maximum and the maximum of the CR-induced atmospheric ionization coincide. At Earth, this maximum is present at altitudes between 16 - 25 km, strongly depending on the geomagnetic field strength \citep[see, e.g.,][]{Herbst-etal-2013}, and on the solar activity \citep[see, e.g.,][]{Bazilevskaya-etal-2008}. At Venus, however, the maximum is located at around 64 km \citep[see, e.g.,][]{Nordheim-etal-2015, Dartnell-etal-2015, Plainaki-etal-2016, herbst2019venus}, which is well within the cloud layer. The reason for the discrepancy of the terrestrial and the Venusian Pfotzer maximum lies in the density of the planetary atmosphere. As shown in Fig.~\ref{fig:1}, where we plot the altitude-dependent temperature profiles of Earth (in blue) and Venus (in black), the surface pressure of Venus is about 90 times higher, and the atmosphere is therefore much more massive than that of its terrestrial sibling. Fig.~\ref{fig:1} also shows the Venusian cloud layers in shades of orange, and the green band represents the possible habitable zone (HZ) of Venus, which was defined by \citet{Dartnell-etal-2015} to extend between 51 km (338.15 K) and 62 km (253.15 K).

Moreover, the Pioneer Venus Orbiter mission revealed that Venus has no significant global intrinsic magnetic field \citep{Russell-etal-1980, Phillips-etal-1987} that may act as a shield against low- and high-energy charged particles. Nevertheless, because solar UV radiation ionizes neutrals in the upper Venusian atmosphere, an ionosphere is generated. Through the interaction of the solar wind with this induced ionosphere, an induced magnetic field is present, which, according to \citet{Taylor-etal-2018}, is strong enough to minimize the erosion of the upper atmosphere. CRs, however, are far too energetic to be deflected by this weak induced magnetic field. Therefore, like at the Earth's magnetic poles, most of the primary CRs have unrestricted access to the Venusian atmosphere. In addition, Venus is located at a distance of only 0.72~AU from the Sun. Thus, it is exposed to much higher particle fluxes from SEP events. Because of the dense Venusian atmosphere, secondary particle showers will develop more extensively than in the terrestrial atmosphere. On the other hand, only a small fraction of secondary particles will reach the Venusian surface.
As discussed for example by \citet{Nordheim-etal-2015} and \citet{herbst2019venus}, the main driver for atmospheric ion-electron pair production at large depths are CRs. 

Numerical studies by \citet{Grinspoon-Bullock-2007} and \citet{Way-etal-2016} suggested that for at least the first 750 million years, the Venusian atmosphere provided habitable conditions, and that liquid water on the surface was present for nearly two billion years \citep[see, e.g.,][]{Donahue-etal-1982, Donahue-Hodges-1992}. This could have been long enough for life (as we know it from Earth) to have been evolved. As suggested by \citet{Grinspoon-Bullock-2007}, with increasing atmospheric temperature, the surface water may have evaporated, which may have forced any putative microorganisms to migrate into the still habitable Venusian cloud layer. Consequently, these microorganisms may have been able to adapt to the limited-water and low-pH environments of the Venusian clouds \citep[see, e.g.,][]{Schulze-Makuch-etal-2004, Grinspoon-Bullock-2007}. According to \citet{Limaye-etal-2018}, the atmospheric movement of life-essential nutrients, water, and organics is, among others, most likely regulated by diurnal cycles and atmospheric storms. The VeGa 1 and VeGa 2 balloon measurements, for example, showed that the typical vertical motion within the Venusian atmosphere is in the order of 1 to 2 m/s \citep[][]{Blamont-etal-1986}. Thus, Venusian cloud particles are assumed to have atmospheric residence times comparable with the two- to three-month terrestrial Hadley circulation. Based on recent measurements of stationary gravity waves at the top of the Venusian cloud layer, which most likely are the result of the surface topography \citep[see][]{Fukuhara-etal-2017}, it becomes evident that vertical motions within the cloud layers are also possible. Because of the stable Venusian cloud lapse rates, therefore, airborne (nutrient-rich) particles could be sustained within the cloud layer for extended time intervals \citep{Limaye-etal-2018}. \citet{Dartnell-etal-2015} investigated whether GCRs and SEPs, including extreme SEP events from the historical record, could represent a constraint on the possibility of cloud-based life on Venus. This study, however, did not include modeling on water-based shape-, size-, and composition-mimicking detectors (phantoms), which more realistically describes the radiation effects in living organisms.

In this study, we revisit the altitude-dependent Venusian radiation environment by studying the cosmic-ray induced atmospheric absorbed and effective dose rates. To study the impact of CRs from the astrobiological point of view, for the first time we study the radiation dose that different phantoms, in particular, a microbial cell and a phantom mimicking human tissue (based on the International Commission on Radiation Protection (ICRP) phantom \citep[see, e.g.,][]{ICRP-2007} would suffer within the Venusian atmosphere. The cosmic-ray induced radiation dose is modeled with the newly developed Geant4-based Atmospheric Radiation Interaction Simulator \citep[AtRIS,][]{banjac-etal-2018}, a 3D simulation code developed to model the radiation dose of exoplanets, which varies from hot Jupiters to Earth-like planets. In addition to the GCR-induced atmospheric radiation due to primary particles with  $Z = 1 - 28$, we investigate the influence of 67 recent GLE events that occurred on Earth between 1942 and 2017 \citep[see, e.g.,][]{Raukunen-etal-2018}. From historical records of so-called cosmogenic radionuclides, it is moreover known that much stronger SEP events occurred in the past \citep[see, e.g.,][]{Mekhaldi-etal-2015}. Two events therefore are of particular interest: the Carrington event, and the AD775 event. Based on assumptions about the GLE event spectrum \citep[see][]{Dartnell-etal-2015}, we also investigate the influence of such intense events on the Venusian atmospheric radiation and their influence on the habitability within the Venusian HZ in the cloud layer.
%
\section{Atmospheric Radiation Interaction Simulator (AtRIS)}\label{sec:approach}
AtRIS, a Geant4-based (Monte Carlo) particle transport code, was recently developed by \citet{banjac-etal-2018} in order to study, among others, the propagation and interaction of energetic particles within diverse (exo)planetary atmospheres and surfaces. As particular input, the user has to define an interface for the planetary atmospheric model. By now, interfaces for the terrestrial NRLMSISE-00 \citep{picone2002nrlmsise} model, the Mars Climate Database (MCD), and the day-side model of the Venus International Reference Atmosphere \citep[VIRA, see, e.g., ][]{Kliore-etal-1985} have been implemented. Thereby, atmospheric altitude-dependent ion and electron pair production rates, secondary particle distributions, and absorbed dose rates and dose equivalent rates can be calculated. Thus, it is rather easy to implement GCR and SEP event spectra. Furthermore, the user can decide on any hadronic and electromagnetic physics lists provided through the standard Geant4 message and compounded physics list naming scheme \citep{Allison-etal-2016,collab2017prm}. 

The main features of AtRIS are i) the planet specification format (PSF), ii) the atmospheric response matrices (ARMs), which quantify the relationship between primary energy and altitude-dependent ionization, absorbed dose rate and dose equivalent, and  iii) the spectrum-folding procedure used to calculate net quantities such as the electron-ion pair production rate by implementing a convolution of a spectrum and the ARM. A detailed description of the main features is given in \citet{banjac-etal-2018}. At this point,  AtRIS has successfully been validated for the terrestrial, the Martian, and the Venusian atmosphere \citep[see][respectively]{banjac-etal-2018, Banjac-etal-2019b, Guo-etal-2019, herbst2019venus}.

\citet{herbst2019venus} calculated the atmospheric ionization within the Venusian atmosphere. To do this, the mean energy required to produce an ion-electron pair had to be taken into account. However, different values exist in the literature: while \citet{Borucki-etal-1982} gave a value of 33.5~eV, \citet{Wedlund-etal-2011} suggested that a value of 28.7 $\pm$ 4.3 eV might be more reasonable in CO$_2$-dominated atmospheres, which leads to differences in the ionization rates in the order of $15\%$ \citep[see][]{herbst2019venus}. These differences are also expected when the atmospheric radiation dose is modeled. However, for comparative reasons, all results discussed in this study are based on the "traditional" value proposed by \citet{Borucki-etal-1982}. Nevertheless, as previously discussed in \citet{Nordheim-etal-2015} and \citet{herbst2019venus}, dividing the presented ionization and radiation profiles by a factor of 1.17  results in values corresponding to the use of the ionization potential by \citet{Wedlund-etal-2011}.

One of the main AtRIS outputs is the so-called ARMs, which reflect the altitude-dependent ionization, absorbed dose, and dose rate equivalent as a function of the primary particle energy. According to this, the \textit{\textup{absorbed dose}} of a specific phantom is derived using the following approach:
\begin{enumerate}
    \item Definition of the particle type $j$ and its kinetic energy E$_i$.
    \item Using the AtRIS-specific precalculated \emph{\textup{relative ionization efficiency}} $\mathcal{I}_{R,j}$ , which is given by
\begin{equation}
    \mathcal{I}_{R,j} \left(E_i\right):= \frac{E_d}{E_i},
\end{equation}
where $E_{d}$ represents the average ionization energy a particle causes within a well-defined phantom.
\item Calculation of the average absorbed dose a phantom experiences when particles of a given energy pass through it. We note that the absorbed dose corresponds to the ionization energy per mass of the phantom $m(r) = \rho \cdot \frac{4}{3}\pi \cdot r^3$. Thus,
\begin{equation}
        \mean{D_j}\left(E_i, r\right) = \mathcal{I}_{R,j}(E_i) \cdot \frac{E_i}{m(r)}
    .\end{equation}
\item If required, multiplication with the geometric factor $g = \pi A(r) = 4\pi^2 \cdot r^2$ \citep[][]{sullivan1967geometric}, which is proportional to the particle flux going through the phantom, represents the normalized average dose $\mean{D_{j, n}}$,
\begin{align}
        \mean{D_j}\left(E_i, r\right) &= \mathcal{I}_{R,j}(E_i) \cdot \frac{E_i}{m(r)} \cdot g \nonumber\\ 
        &= \mathcal{I}_{R,j}(E_i) \cdot E_i \frac{3}{\rho \cdot r}
\end{align}
We like to point out that the shape and size of different phantoms can vary. Different shapes, however, account for different geometric factors: $g$ = $\pi \cdot 2 a^2$ for a slab detector, $g$ = $\pi \cdot 2 r^2$ for a disk detector, and a spherical detector, for example, a planetary atmospheric layer, has a geometrical factor of $\pi \cdot 4 \pi r^2$ \citep[see, e.g.,][]{Banjac-etal-2019b}. By placing phantoms of different sizes and shapes within the same radiation field, the number of particles that can penetrate the phantom and thereby deposit energy within, therefore decreases as the square of the phantom radius.
\item Convolution with the primary particle spectrum and summation over all energy bins and particles.
\end{enumerate} 
In this way, the dosimetric quantities of a reference phantom can be calculated without the need of further time-consuming computations. This is helpful for any parametric study like the one presented in Sec.~\ref{sec:gles}, where the altitude-dependent absorbed dose rate profiles for 67 GLE events are computed.

The Venusian absorbed dose rates have currently only been computed for a Venusian air phantom \citep[see, e.g.][]{Dartnell-etal-2015}. However, life as we know it is based on water rather than CO$_2$. According to climate modeling efforts of \citet{Way-etal-2016}, Venus may early in its planetary evolution have had a liquid-water ocean and surface temperatures that may have allowed for habitable conditions. This and the fact that with increasing temperatures the Venusian water content may have evaporated and, therewith, microorganisms may have been transported in the cloud layer \citep{Grinspoon-Bullock-2007} lead us to study the absorbed dose in H$_2$O-based phantoms to better describe the impact of CRs in the Venusian atmosphere on life as we know it (see Sec. \ref{sec:phantom}).

We also further calculate the \textit{\textup{dose equivalent}} $H$, which is more relevant for estimating the radiation hazard to H$_2$O-based (human-like) phantoms because it takes into account the varying biological effectiveness of different radiation types. With this, the dose equivalent $H$ is calculated by
\begin{equation}
    H=\sum_{j}W_{R,j}\cdot D_{j},
    \label{eq:equi}
\end{equation}
where $w_{R,j}$ is the radiation weighting factor defined by \citetalias{protection1991icrp} and most recently modified in \citetalias{valentin2007icrp}. Thereby, $w_{R,j}$ account for the different biological effectiveness of different types of radiation and for all particles $w_{R,j}\ge 1$, as listed in Table~\ref{tab:wj}.
%
\begin{table}[]
\begin{center}
\caption{Weighting factors $w_{R,j}$ defined by \citet{Petoussi-Henss-etal-2010}.}
\label{tab:wj}
\begin{tabular}{l l}
\hline
\hline
Radiation type $j$ & $w_{R,j}$ \\
\hline
Photons, electrons, muons & 1 \\
protons, charged pions & 2\\
$\alpha$, fission fragments, heavy ions & 20\\
neutrons & see Eq.~(\ref{eq:rwf})\\
\hline
\end{tabular}
\end{center}
\end{table}
%
For neutrons, \citetalias{valentin2007icrp} defines a continuous function given by
\begin{equation}
    w_{R,n}=\begin{cases}
         2.5 + 18.2\cdot e^{\frac{-\left[ \ln\left(E\right)\right]^2}{6} },&\text{for } E < 1\,\text{MeV}  \\ 
         5.0 + 17.0\cdot e^{\frac{-\left[ \ln\left(2\cdot E\right)\right]^2}{6} },&\text{for } E\in\left[1\,\text{MeV}\--50\,\text{MeV}\right]  \\ 
         2.5 + 3.25\cdot e^{\frac{-\left[ \ln\left(0.04\cdot E\right)\right]^2}{6} },&\text{for } E> 50\,\text{MeV} 
    \end{cases}
    \label{eq:rwf}
,\end{equation}
where $E$ is the kinetic energy of the neutron in MeV. For all other particles, $w_R$ is equal to $1$. Because $w_R$ is dimensionless, the dose equivalent $H$ has the same dimension as the absorbed dose. A more advanced approximation of $w_{R,j}$ exists in the literature. However, this approximation relies on the determination of the linear energy transfer (LET), which is difficult and somewhat complex. Although the approximation can correct for overestimating the influence of relativistic heavy minimally ionizing particles (MIPS), for example, 100 GeV $\alpha$ particles, it is not suitable for cases in which the LET spectrum cannot be measured. Moreovoer, at column depths below 200 g/cm$^{2}$ (corresponding to an altitude of $\sim$ 62.5 km within the Venusian atmosphere), the flux of relativistic heavy particles is relatively low, so that Eq.~(\ref{eq:rwf}) is a valid approximation. Furthermore, we note that these weighting factors are tailored explicitly for biological studies with a focus on human radiation health. Thus, they may not be directly applicable to astrobiology studies with a focus on the survival of microorganisms, for example.
%
\section{Simulation setup}
As discussed in \citet{herbst2019venus}, the Venusian environment was modeled for spherical geometry using a core with a radius of $\sim$ 6052 km, a soil composition of 50\% Si, 40\% O, and 10\% Fe, and a crust (soil) sheet with a thickness of 100~m. The upper layer of the Venusian atmosphere was set at an altitude of 150 km, and the atmosphere was divided into 500~m thick layers. Furthermore, the same atmospheric composition, source altitude, and particle input spectra were used.

The atmospheric setup was thus based on the VIRA model \citep[see][]{Kliore-etal-1985}, which uses the day-side atmospheric parameter set by \citet{Seiff-etal-1985} at low- to mid-atmospheric layers (surface to 100 km) for low latitudes ($\phi<30^{\circ}$) and those proposed by \citet{Keating-etal-1985} for altitudes up to 150 km at low latitudes ($\phi<16^{\circ}$). The atmospheric composition was chosen to consist of 96.5\% CO$_2$ and 3.5\% N$_2$. 

To simulate the hadronic and electromagnetic interactions, and thus the evolution of secondary particle cascades as well as the atmospheric radiation dose, \citet{Dartnell-etal-2015} based their studies on the QGSP$\_$BIC$\_$HP model using the quark-gluon-string (QGS) model for high-energy particle interactions and the binary cascade model (BIC) for particle energies below 10 GeV. As discussed in \citet{herbst2019venus}, the more accurate model to use is the FTFP$\_$BERT$\_$HP model, which is explicitly recommended for radiation protection and shielding applications. Therefore, our computations use the Bertini-style cascade for hadrons with energies below 5 GeV and the Fritiof (FTF) model \citep[see, e.g.,][]{Nilsson-Almqvist-Stenlund-1987} for particle interactions of mesons, nucleons, and hyperons with energies between 3 GeV and 100 TeV. In addition, the standard em constructor \citep[see, e.g.,][]{Allison-etal-2016} was used to model electromagnetic interactions. We note that differences due to the use of these two different hadronic interaction models only occur at altitudes above 120 km and below 5 km \citep[see Fig. 4 in][for more details]{herbst2019venus}. Therefore, in the context of the habitability of the Venusian cloud layer, the computed physical radiation doses of our study are comparable to those of \citet{Dartnell-etal-2015}.

As discussed in \citet{herbst2019venus}, we used the CREME2009 model\footnote {available at \url{https://creme.isde.vanderbilt.edu/}} \citep[see, e.g.][]{Tylka-etal-1997} as input for the time-dependent GCR flux at the top of the Venusian atmosphere. This model provides the differential intensities for all particles from protons (Z = 1) to nickel (Z = 28) for energies between 1 MeV/nuc up to 100 GeV/nuc derived for a solar distance of 1 AU. In this study, we investigate the atmospheric radiation dose during quiet solar minimum and maximum conditions. As discussed in \citet{herbst2019venus}, the high-energy tail of the particle spectrum (in particular up to 10 TeV/nuc) may have a much higher impact on the atmospheric radiation dose as previously thought. Therefore, an element-dependent power-law extension of the spectra was applied. Although Venus is about 30$\%$ closer to the Sun than Earth, a rescaling of the modeled GCR spectra is not necessary because the radial gradients of GCR particles in the inner Solar System are negligible \citep[see, e.g.,][]{Heber-etal-1996, Morales-Olivares-Caballero-Lopez-2010, Gieseler-Heber-2016}. 

Furthermore, the spectra of particularly strong SEP events are provided by the CREME2009 model. Because of their acceleration at shock fronts of solar flares (SFs) and coronal mass ejections (CMEs), SEPs have energies of only up to some hundred MeV and reach the Solar System planets anisotropically. In contrast to GCRs, the flux of SEP events strongly depends on the orbital distance. Thus, a 1/R$^2$ scaling was applied to the spectra. Following the investigations by \citet{herbst2019venus}, we studied the influence of 67 GLE events measured within the neutron monitor era \citep{Raukunen-etal-2018}. The strongest GLE event ever measured in situ occurred on 23 February 1956 (hereafter GLE05). However, based on observations of the so-called Carrington event, and from the cosmogenic radionuclide records of $^{10}$Be, $^{14}$C, and $^{36}$Cl covering the past $\sim$ 10,000 years, today we now know that even much stronger SEP events have occurred throughout the Sun's history, in particular around BC 660, AD774-775, and AD993-994 \citep[see][respectively]{OHare-etal-2019,  Mekhaldi-etal-2015, Miyake-etal-2012}. Based on model efforts, \citet{Kovaltsov-etal-2014} and \citet{Herbst-etal-2015} found that the event of AD775 was most likely well over one order of magnitude stronger than GLE05.

We here investigate the impact of different GCRs nuclei, their upper energy limit, and the impact of solar modulation on the Venusian atmospheric radiation dose, and we highlight the differences between absorbed dose and dose equivalent. We also study the influence of 67 terrestrial GLE events on both quantities and discuss the impact of potential super-events like the Carrington- and the AD774/775 event. However, because the spectral shape and the upper event energy of the latter is not known, the influence of the two historical events was modeled based on measured event spectra of other historical events.
\section{Results and discussion}
\begin{figure*}[!t]
\centering
\includegraphics[width=\textwidth]{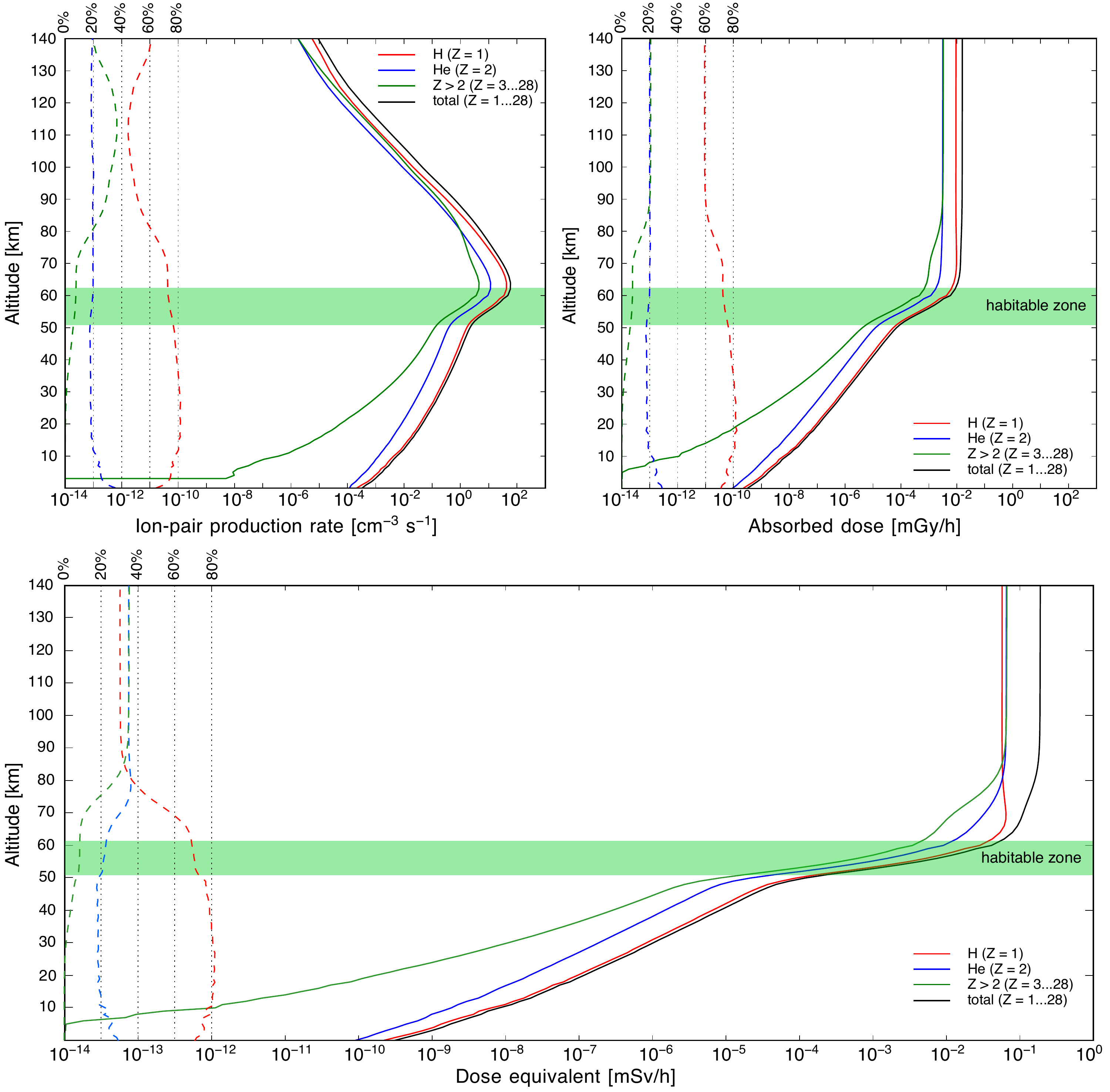}
\caption{Left upper panel:Altitude-dependent ion pair-production rates induced by primary protons (Z=1, solid red line), alpha particles (Z=2, solid blue line), and Z=3-28 nuclei (solid green line). The total ion-pair production is given as solid black line. Right upper panel: Modeled GCR-induced altitude-dependent absorbed dose rates. Lower panel: Corresponding dose-equivalent rates. All three panels show the relative contributions of the GCR nuclei as dashed lines (notation on the upper x-axis).}
\label{fig:2}
\end{figure*}
\subsection{Contribution of different GCR nuclei on the atmospheric ionization}
As shown in \citet{herbst2019venus}, the atmosphere primarily is ionized through GCR protons (Z=1) and helium nuclei (Z=2). The left panel of Fig.~\ref{fig:2} shows a more detailed study of the particle-dependent GCR-induced Venusian ionization of Z = 1 (solid red line), Z = 2 (solid blue line), and Z = 3-28 (solid green line) particles. Information on the relative contribution of the different GCR species is also given as dashed lines on the left-hand side. The corresponding values are given in the upper x-axis. Primary protons are the main source of the CR-induced ionization at all altitudes, with increasing significance toward lower altitudes (up to 80$\%$). An anticorrelated behavior can be found for Z= 3-28 primary particles. While accounting for about 40$\%$ of the Venusian ionization at altitudes down to 90 km, their input becomes much smaller (below 10$\%$) at altitudes below 70 km, where their input can be neglected. Primary alpha particles, however, have a constant relative contribution of 20$\%$ at altitudes above 10 km. Below this, their influence increases up to 40$\%$; furthermore, a unique Venusian feature is visible: the ionization profile is well within the upper cloud layer (57 km - 70 km), and thus also well within the upper HZ.
\subsection{Contribution of different GCR nuclei to the absorbed dose and the dose equivalent}
By applying the approach discussed in Sec.~\ref{sec:approach}, we modeled the altitude-dependent absorbed dose rates and the equivalent dose rates. We show them for example in Fig.~\ref{fig:2}. Although the atmospheric ionization (left panel) and the absorbed dose rates (right panel) are closely related, considerable differences in the two altitude-dependent profiles are clear. The total absorbed dose rate (solid black line) and the relative contributions of protons (in red), alpha particles (in blue), and Z>2 nuclei (in green) are constant in between 140 km and 85 km. Here, protons account for 60$\%$ of the total absorbed dose, while alpha particles and Z>2 nuclei account for about 20$\%$ each. Below 85 km, this picture changes. Here, the influence of Z>2 nuclei rapidly decreases, while the influence of protons increases up to 80$\%$. However, down to 10 km, the contribution of alpha particles is always within 20$\%$, but it increases to 40 $\%$ at altitudes below. Within the potential Venusian habitable zone (highlighted in green, 51 km - 62 km), for example, protons contribute more than 70$\%$ to the total absorbed dose rate, while Z$>2$ particles contribute less than 10$\%$ and can therefore be neglected.

In order to give a more realistic estimate of the radiation hazard  to  potential  microorganisms  in  the  Venusian  clouds, it  is more convenient to investigate the dose equivalent that accounts for the biological effectiveness of different radiation types (see Section~\ref{sec:qbar_gcr} for more details), as shown in the lower panel of Fig.~\ref{fig:2}. Here the impact of primary particles with Z$>$2 between 140 km and 85 km becomes evident because with a contribution of more than 35$\%$ to the total equivalent dose, they are the dominant source. However, although they have a much higher biological effectiveness than the primary protons and alpha particles, they do not deliver the majority of the equivalent dose at all altitudes. Below 85 km, their contribution clearly rapidly decreases below 10 $\%$ within the potential HZ.
In order to model the absorbed dose rates and the dose equivalent rates from the upper cloud layer down, it is therefore sufficient to take only the interactions of GCR protons and alpha particles into account.
\subsection{Investigating absorbed dose rates in different phantoms}\label{sec:phantom}
\begin{figure}[!t]
\centering
\includegraphics[width=0.7\columnwidth]{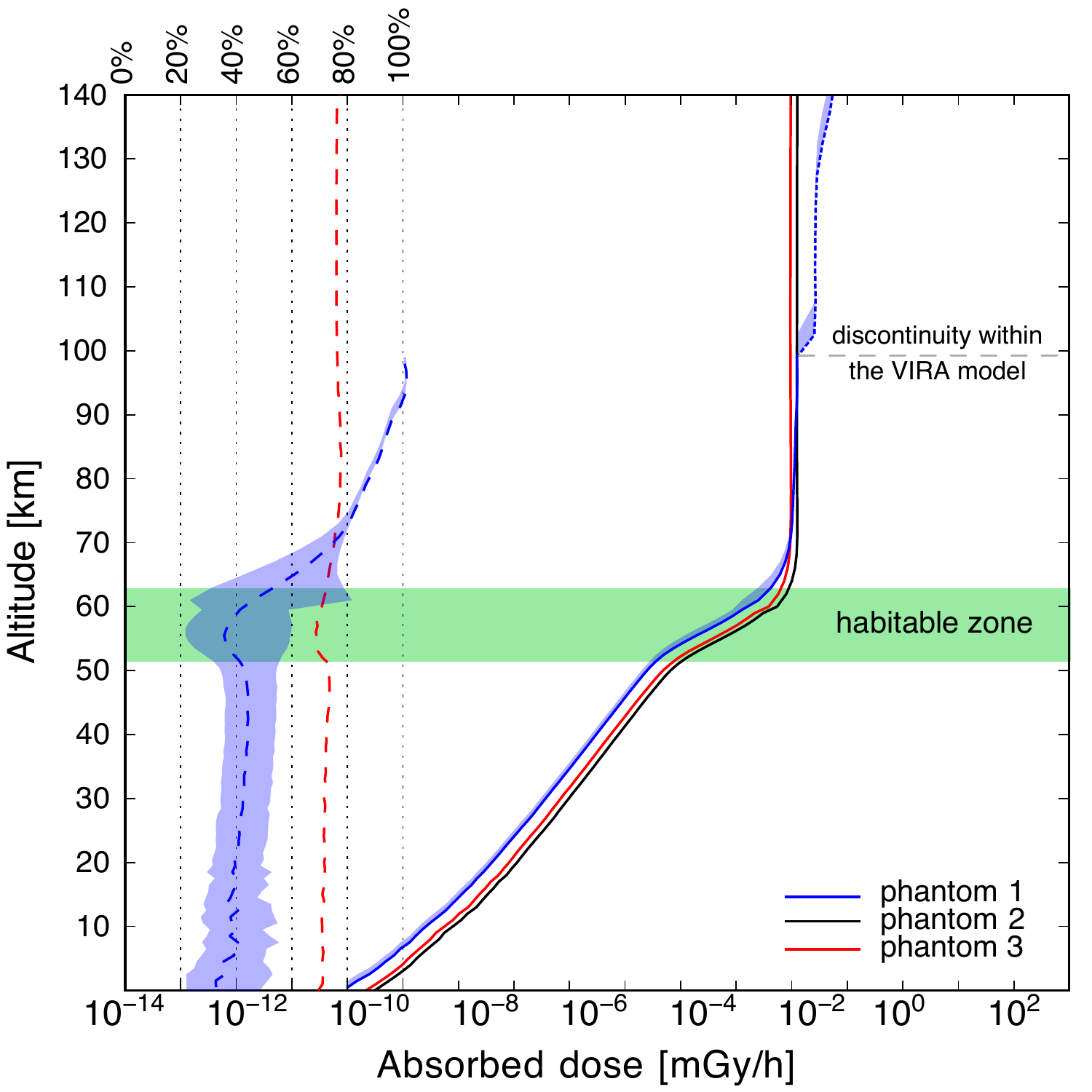}
\caption{Altitude-dependent absorbed dose rate profiles encountered by the (H$_2$O-based) ICRU phantom (solid black line), the Venusian air (solid blue line), and a microbial cell (solid red line). Also shown are the relative differences with respect to the ICRU sphere, displayed as dashed lines. For the dose in the air plot (blue line), the relative difference includes dotted lines that correspond to maximum and minimum differences, as described in detail in the text.}
\label{fig:3}
\end{figure}
\begin{table}[]
\caption{Modeled absorbed dose-rate values at the lower and upper limit of the HZ (51 km and 62 km, respectively) proposed by \citet{Dartnell-etal-2015}. Here, we list the values corresponding to the three scenarios.}
\label{tab:1}
\begin{tabular}{c c c c}
\hline
\hline
                  & \multicolumn{3}{c}{Absorbed dose rates {[}mGy/h{]}}                                  \\
Altitude {[}km{]} & phantom 1          & phantom 2                      & phantom 3                      \\
\hline
51                & 2.83$\times$10$^{-5}$ & 9.82$\times$10$^{-5}$ & 6.87$\times$10$^{-5}$ \\
62                & 2.49$\times$10$^{-3}$ &      8.02$\times$10$^{-3}$               & 5.54$\times$10$^{-3}$ \\
\hline
\end{tabular}
\end{table}
A commonly used approach to determine the (Venusian) atmospheric radiation dose is to divide the calculated ion-electron production rates by the density of the ambient planetary atmosphere \citep[see, e.g., ][]{Dartnell-etal-2015}. In this way, \citet{Dartnell-etal-2015} analyzed the radiation constraints on a potential Venusian aerial biosphere. They  only considered the radiation dose on the gas, however, and therefore did not explicitly study the radiation impact on biological systems such as cells, microorganisms, or organic tissue. 
To study the exposure of life as we know it, phantoms that better reflect biological systems can therefore be used. The  ICRU proposed the so-called ICRU sphere, a tissue-equivalent sphere with a radius of 15 cm and a density of 1 g/cm$^3$. With a composition of 76.2\% oxygen, 11.1\% carbon, 10.1\% hydrogen, and 2.6\% nitrogen, the ICRU phantom is designed to represent and reflect life as we know it from Earth \citep{icru-1980}.

However, because it is not clear if (possible) life on Venus has formed and evolved in comparable prebiotic conditions, using this phantom to model the radiation hazard of Venus may lead to false conclusions. Because life is in general expected to be water based \citep[see, e.g.,][]{Ball-2017}, a more realistic approach needs to be taken in order to give a more advanced estimate of the altitude-dependent radiation hazard to potential Venusian life. In this section, we therefore investigate the impact of CRs on different phantoms:
\begin{itemize}
    \item \textit{phantom 1}: based on Venusian air \citep[as discussed in][]{Dartnell-etal-2015}
    \item \textit{phantom 2}: an ICRU-sphere equivalent consisting of 100$\%$ liquid water with a radius of 15~cm
    \item \textit{phantom 3}: a water-based phantom corresponding to the size of a microbial cell of spherical shape and a radius equal to 50~$\mu$m
\end{itemize}
The GCR-induced absorbed dose rate profiles for phantom 1 (in blue), phantom 2 (in black), and phantom 3 (in red) are shown in Fig.~\ref{fig:3}. Different approaches lead to different estimates of the altitude-dependent radiation dose. Consequently, the modeled absorbed dose rates depend on the size and the composition of the assumed phantom. While the absorbed dose rates above 100 km are much higher for phantom 1, the dose rates below 70 km are up to 60$\%$ lower than the absorbed dose phantom 2 would suffer within the Venusian atmosphere (see the dashed lines for the relative differences with respect to phantom 2). A comparison of the absorbed dose rate values at the upper and lower limit of the HZ is given in Table~\ref{tab:1}. Within the HZ, the absorbed dose rates of phantom 2 (phantom 3) are up to four (three) times higher than those of the Venusian air (phantom 1). We note that, due to its composition, phantom 2 creates an appreciable amount of self-irradiation, for example, due to the production of secondary particles within the 15 cm water-sphere, which further leads to an additional energy deposition. However, compared to phantom 3, the 50$\mu$m  water-based microbial cell, this effect only accounts for differences in the order of a factor of ~ 1.7 and does not explain the differences, in particular, those between phantom 1 and phantom 3.

Consequently, the previously introduced approach by \citet{Dartnell-etal-2015} systematically underestimates the absorbed dose of potential water-based microorganisms below 70 km, and in particular, within the potential Venusian HZ, by up to 45$\%$. 

We note that the absorbed dose rates of phantom 2 have a systematical error at altitudes above $\sim$ 100~km that results from the use of the VIRA model with the two atmospheric models by \citet{Seiff-etal-1985} and \citet{Keating-etal-1985}, which are discontinuous at their interface. As discussed in Section~\ref{sec:approach}, the absorbed dose rates of phantom 2 were computed by dividing with the atmospheric density $\rho$ of the ambient air, which then results in a jump of the absorbed dose rate values. This jump is not visible in \citet{Dartnell-etal-2015}, therefore we assume that a smooth transition between the two atmospheric models was implemented.

Liquid water may have been present on the Venusian surface. Model studies by \citet{Grinspoon-Bullock-2007} suggest that this water may have evaporated and in addition, transported microorganisms into the Venusian cloud layer. Our further investigations therefore use the water-based 50$\mu$m microbial cell phantom (if not stated otherwise).
\subsection{Investigating the influence of the upper CR energy limit}
\citet{herbst2019venus} showed that it is crucial to consider primary protons up to energies of 10~TeV in order to accurately reflect the Venusian atmospheric ion-pair production rate at altitudes below 40~km. In Fig.~\ref{fig:4} we follow up on this investigation by studying the altitude profile of the induced absorbed dose rates as a function of the maximum primary proton energy. Although substantial differences between the upper limit of 100 GeV (solid blue line), 1 TeV (solid red line), and 10 TeV (solid black line) can be observed at altitudes below 40 km, only differences of up to 20 $\%$ can be found between an upper limit of 100 GeV and 10 TeV within the potential HZ. While studies of the radiation dose profiles of the entire altitude should include primaries with energies up to 10 TeV, this is therefore not mandatory when the habitability of a possible aerial biosphere is investigated. An upper limit of 1~TeV, as assumed by \citet{Dartnell-etal-2015}, is sufficient here.
\begin{figure}[!t]
\centering
\includegraphics[width=0.7\columnwidth]{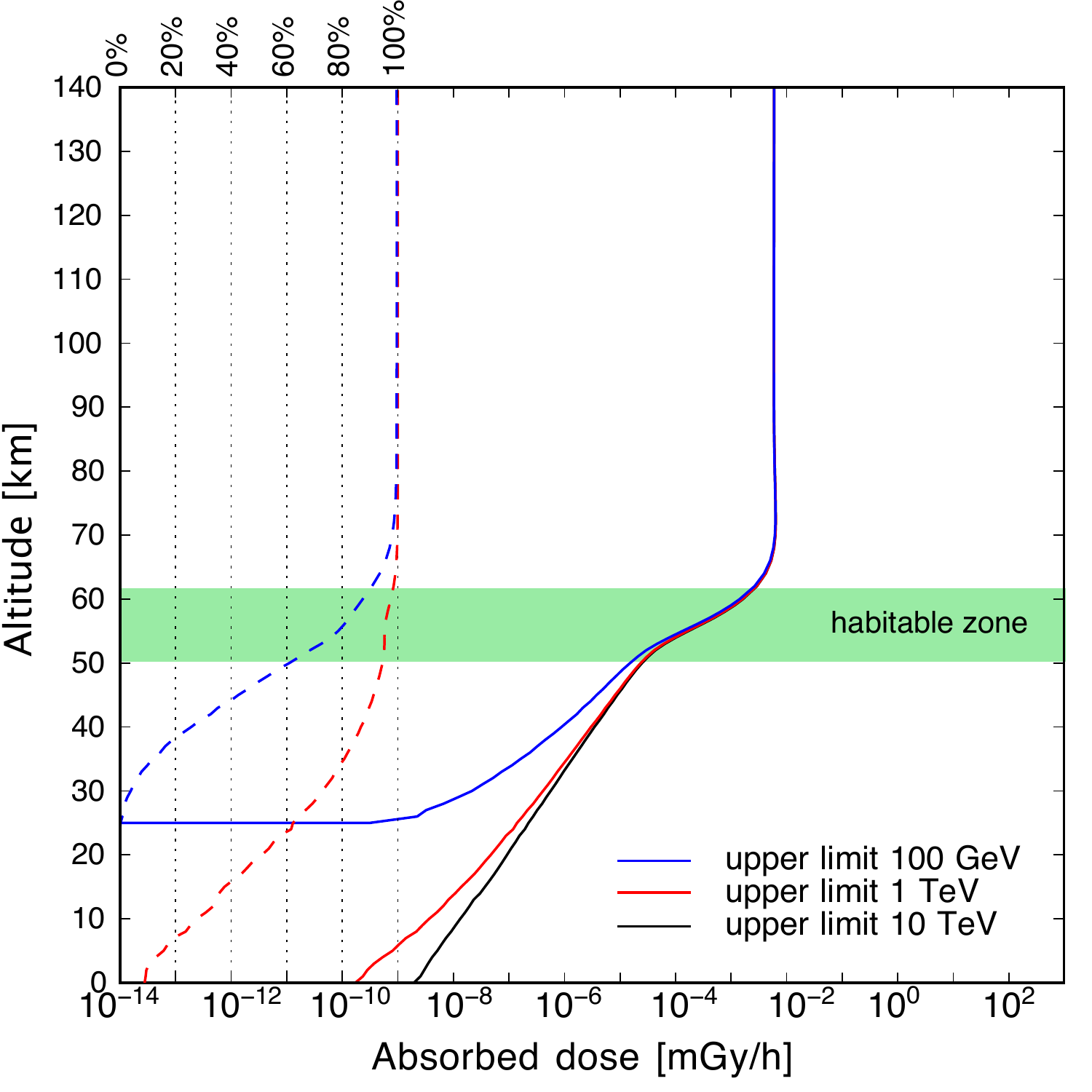}
\caption{Altitude-dependent absorbed dose rate profiles depending on the upper energy limit. The solid blue, red, and black lines correspond to the upper limits equal to 100~GeV, 1~TeV, and 10~TeV. The dashed lines on the left indicate the relative difference with respect to the results based on an upper energy limit of 10~TeV.}
\label{fig:4}
\end{figure}
\subsection{Investigating the influence of solar activity}
We used the CREME model by \citet{Tylka-etal-1997} to calculate the primary particle spectra of Z= 1-28 nuclei as a function of the solar modulation. In order to study the influence of the modulation of GCR particles upon the Venusian atmospheric radiation dose, the absorbed dose rates during solar minimum and maximum conditions were computed. As shown in Fig.~\ref{fig:5}, the altitude-dependent values during low solar activity are given in black, and those during solar maximum conditions are displayed in red. 

The relative differences on the left show with respect to the induced radiation doses during solar minimum conditions (dashed lines) that the modulation of GCRs due to the solar activity only affects altitudes above 50 km and also the potential HZ. While the dose rates induced during high solar activity are about 50 $\%$ lower than the absorbed dose during solar minimum at altitudes above 80~km, the differences within the Venusian HZ still account for up to 20$\%$ and therefore are about 1.4 times higher during solar minimum conditions.
\begin{figure}[!t]
\centering
\includegraphics[width=0.7\columnwidth]{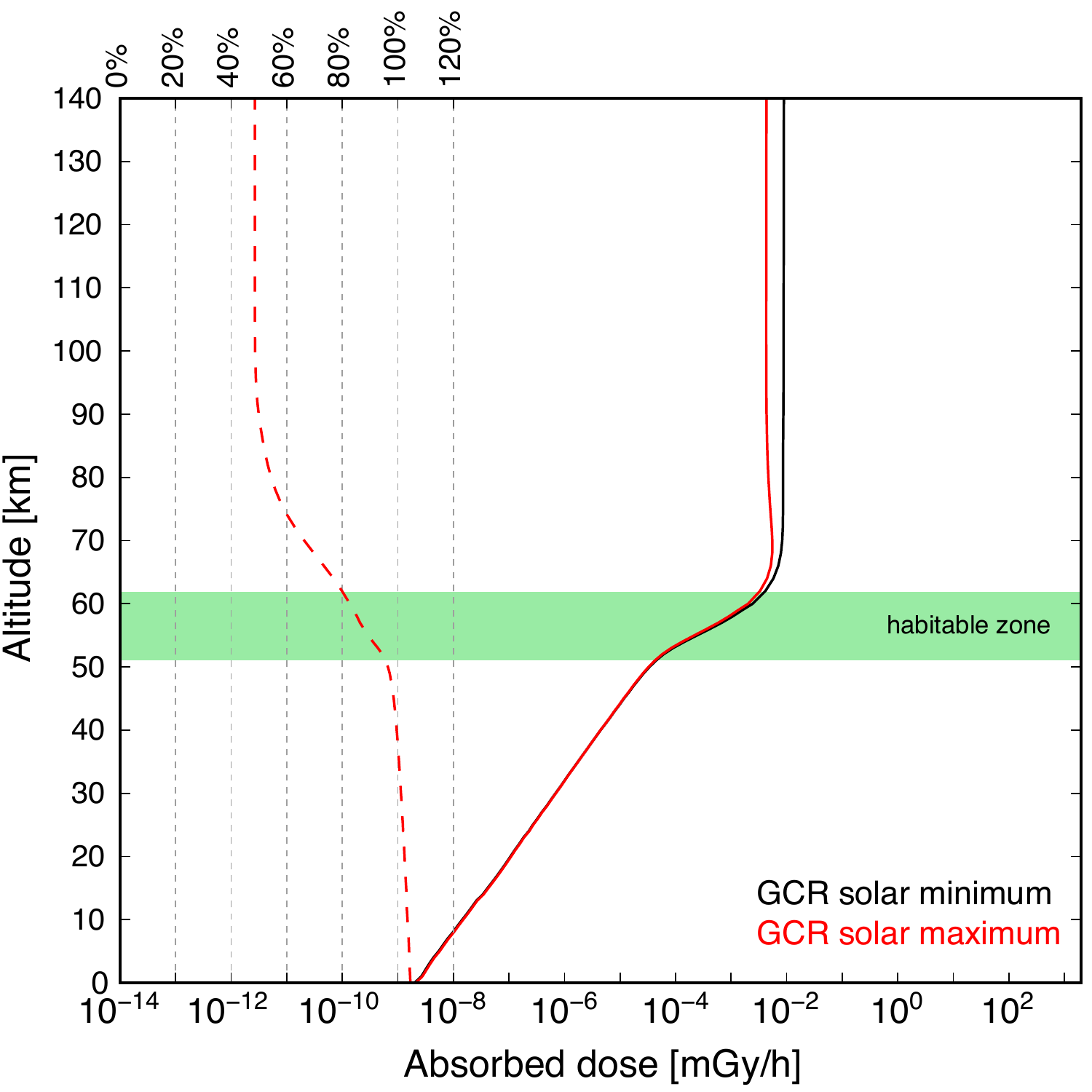}
\caption{Comparison of the absorbed dose rates during solar minimum (solid black line) and  solar maximum (solid red line), The relative difference between both quantities is shown as the red dashed line on the left.}
\label{fig:5}
\end{figure}
\subsection{Investigating the dose equivalent}\label{sec:qbar_gcr}
According to the approach we discussed in Section~\ref{sec:approach}, the deposited ionization energy of each Z = 1-28 nuclei was scaled with the corresponding radiation weighting factor $w_{R,j}$ given in Table~\ref{tab:wj} and Eq.~(\ref{eq:rwf}). Fig.~\ref{fig:6} shows the corresponding modeled altitude-dependent absorbed dose rate profile (solid black line) and the dose equivalent profile (solid red line) that is the result of  primary protons and helium nuclei alone during solar minimum conditions. 
\begin{figure}[!t]
\centering
\includegraphics[width=0.7\columnwidth]{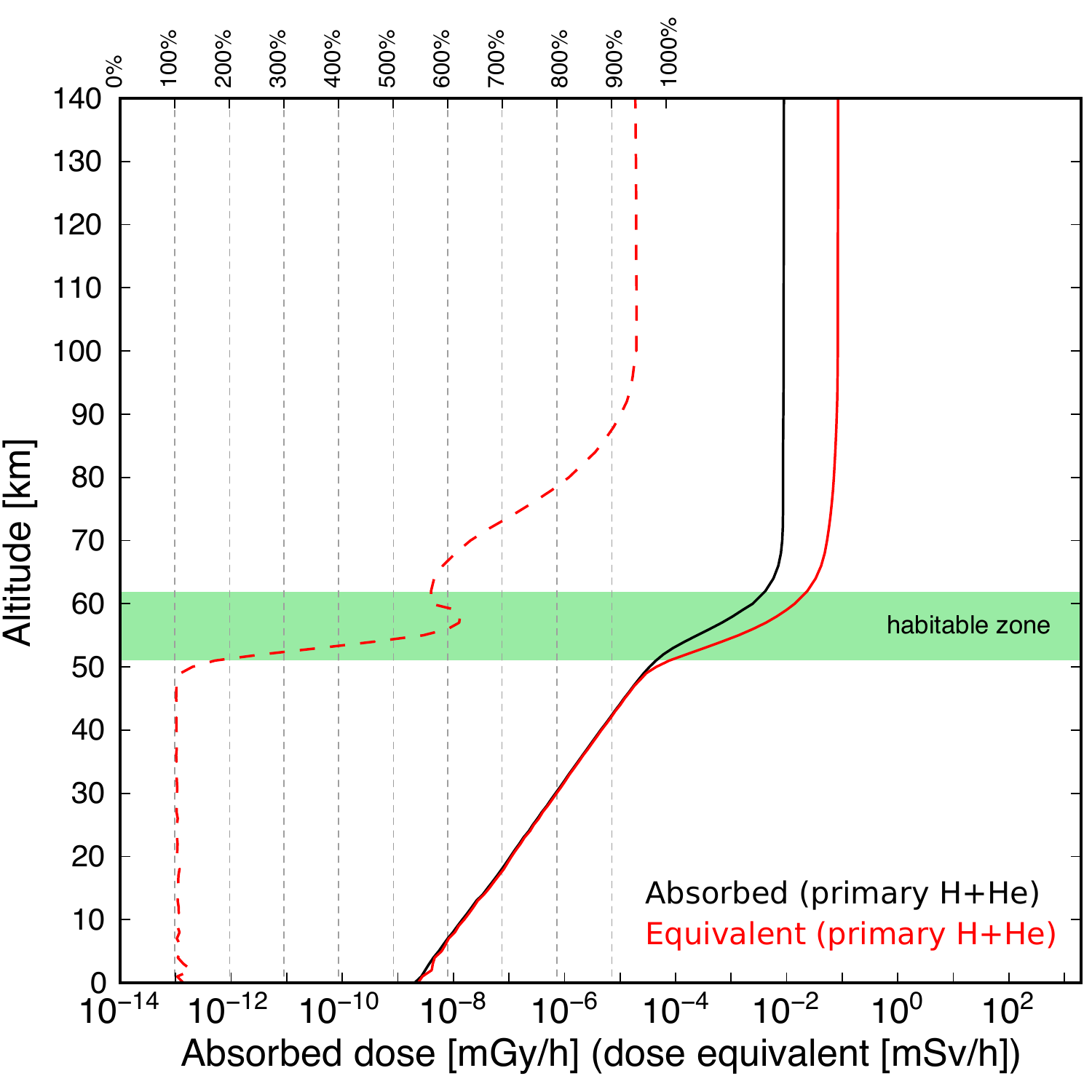}
\caption{Dose equivalent (solid red line), absorbed dose (solid black line), and the obtained average quality factor $\bar{Q}$ (red dashed line). See text for details.}
\label{fig:6}
\end{figure}

As discussed in Sect. 4.1, this is a valid approximation because Z$>$2 nuclei only contribute with less than 20$\%$ to the Venusian radiation dose. The altitude-dependent ratio between the two measures therefore gives us the mean radiation weighting factor, which is comparable to the average quality factor $\mean{Q}$ \citep{protection1991icrp}. At the Martian surface, for example, the average quality factor is around three \citep[see][]{Hassler-etal-2014}.

According to our results, the average quality factor at the upper Venusian cloud layer ($\sim$72 km) is almost ten and slightly decreases toward a second peak within the HZ at around 57 km. At the lower boundary of the HZ and the lower cloud layer, in particular between 57 km and 47 km,  $\mean{Q}$ rapidly drops to negligible values. This drop-off indicates that neither neutrons nor protons or nuclear fragments with the highest $w_{R,j}$ values contribute to the absorbed dose rates below the Venusian cloud layers. We therefore conclude that the atmospheric ionization and the absorbed dose rates at lower altitudes are solely induced by secondary muons.
\subsection{Investigating the solar cosmic-ray-induced atmospheric radiation}
From observations, we know that the spectral shape of GLE events and GCRs differs strongly. Although the energy spectrum of such strong SEPs is orders of magnitudes higher in the low-energy range, their energy spectra often do not exceed energies of a few GeV/nuc. In addition, different particle acceleration and transport mechanisms exist for the two populations. This means that the energy spectra of  GLE events and their temporal evolution differ significantly from one another. 

Although remarkable efforts revealing the temporal evolution of SEP events can be found in the literature \citep[see, e.g.,][and references therein]{Belov-etal-2005, Bombardieri-etal-2007, Plainaki-etal-2007, Matthiae-etal-2009, Miroshnichenko-2018}, the model efforts to study the GLE-induced atmospheric effects are often based on a mean energy spectrum. As previously discussed in \citet{Nordheim-etal-2015}, \citet{Plainaki-etal-2016}, and \citet{herbst2019venus}, the Venusian atmospheric ionization is significantly affected by these strong SEP events. We investigated their influence on the atmospheric absorbed dose rates.

In order to study the mean influence of strong SEP events on the atmospheric Venusian radiation dose, in a first step, we investigated one of the strongest GLE events of the modern era that occurred in October 1989. Based on the studies by \citet{Dartnell-etal-2015}, \citet{Nordheim-etal-2015}, and \citet{herbst2019venus}, the starting time was set at 1300 UTC on 19 October 1989, and two specific cases were investigated: 
\begin{itemize}
    \item \textit{scenario 1}, in which the 5-minute average event peak flux based on the GOES measurements was taken into account, and
    \item \textit{scenario 2}, in which the 180-hour average of the particle flux was used.
\end{itemize}
The results are displayed in the upper panel of Fig.~\ref{fig:7}. To convey an idea of  the impact strength of the event on the Venusian atmosphere, we also plot the GCR-induced radiation dose taking into account primary protons and alpha particles during solar minimum conditions. Even though most GLE events are limited to energies below 10~GeV, strong SEP events like the October 1989 event can significantly increase the Venusian radiation dose rate at altitudes above $\sim$ 50~km, and thus also well within the potential HZ.
\begin{figure}[!t]
\centering
\includegraphics[width=0.5\columnwidth]{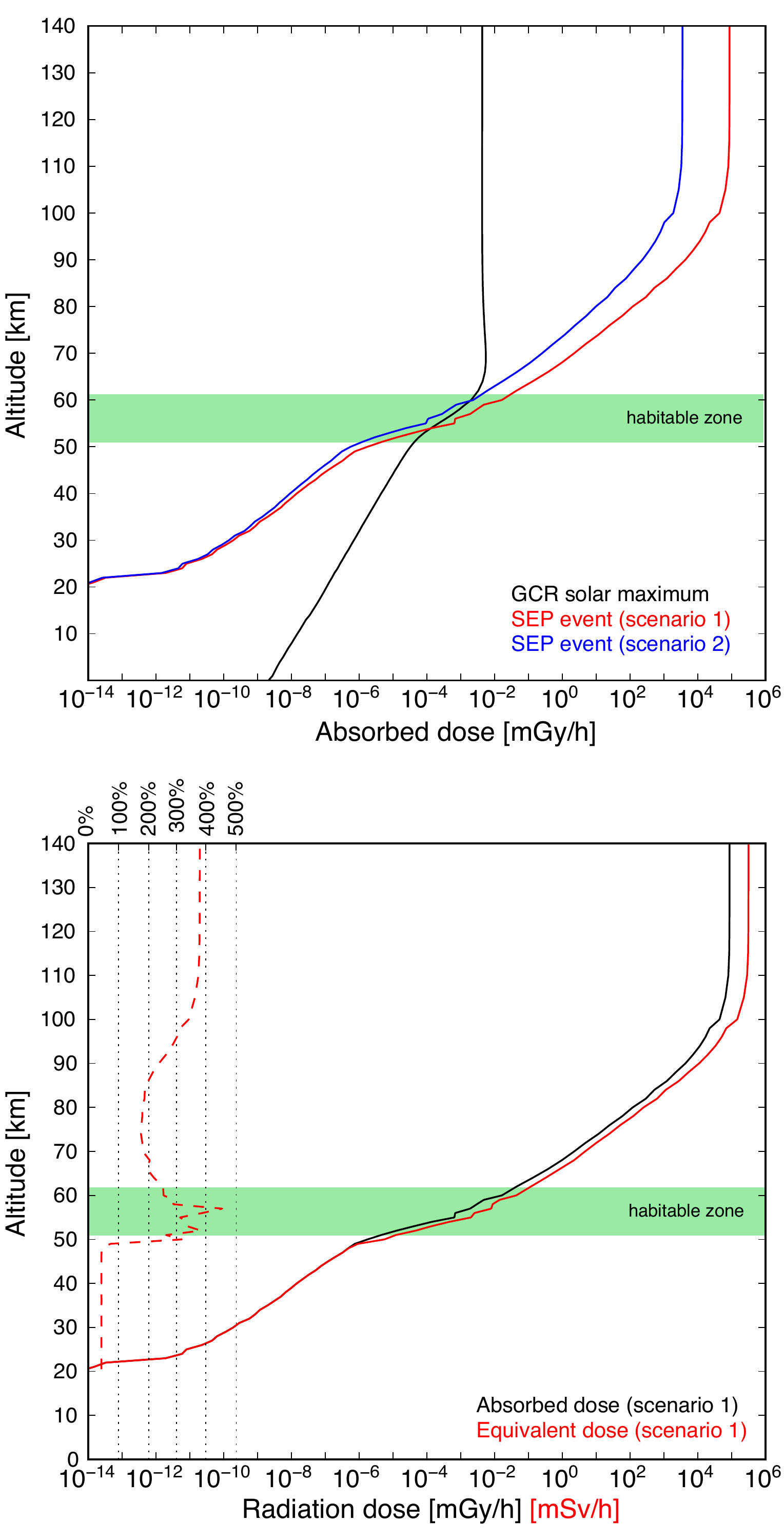}
\caption{Upper panel: Altitude-dependent absorbed dose rates induced by the October 1989 event. The results of scenario 1 are displayed as solid red line, those of scenario 2 as solid blue line. Lower panel: Modeled absorbed dose rate and the dose equivalent of scenario 1. The ratio between the two quantities is given as red dashed line.}
\label{fig:7}
\end{figure}

The hazards of GLE events can better be estimated by studying the dose equivalent altitude profile, which is shown for the example of scenario 2 in the lower panel.  The ratio between the two measures, again reflecting the average quality factor $\mean{Q}$, is plotted on the left. While GCR particles induce a radiation field with $\mean{Q}\approx 10$ (see discussion in Section~\ref{sec:qbar_gcr} and Fig.~\ref{fig:6}) at the top of the cloud layer, the 180-hour average of the GLE in October 1989 shows an average quality factor of $\approx20$, indicating that mostly neutrons are produced within the atmospheric layer. This shows that extreme SEP events can cause significant short-term increases in the radiation hazard in the cloud region above 50~km. 

However, other differences with respect to the GCR-induced radiation dose become apparent: in the case of the GLE-induced radiation field, $\bar{Q}$ is drastically reduced below 50~km. Furthermore, below 20~km, no more secondary particles are produced, while $\mean{Q}$ is below 4 at altitudes above 80~km, which is much smaller than the GCR-induced $\bar{Q}$, as we show in the lower panel of Fig.~\ref{fig:7}.
\begin{figure*}[!t]
\centering
\includegraphics[width=\textwidth]{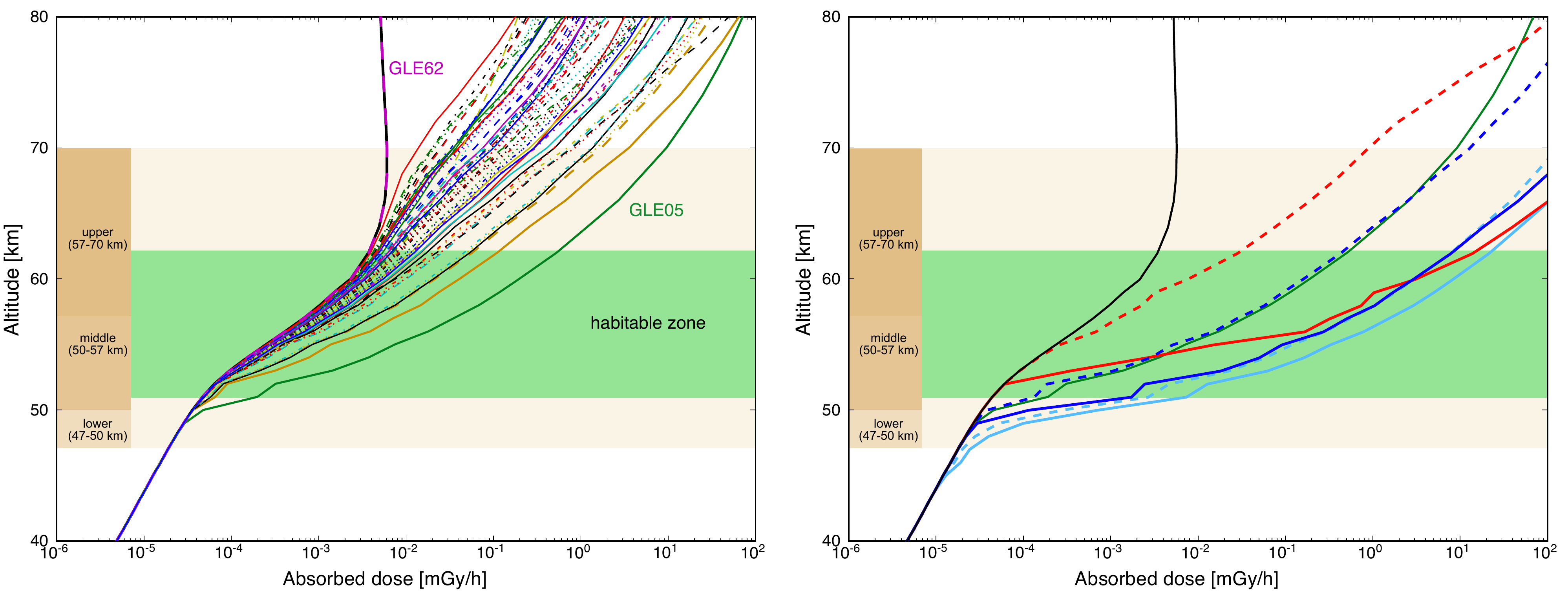}
\caption{Left panel: Influence of modern ordinary strong SEP events on the total (GCR + GLE-induced) Venusian absorbed dose of a water-based microbial cell (colored lines) within the Venusian cloud layers, and thus the potentially HZ. Of particular interest is the influence of the strongest event ever measured: GLE05 (green line). Right panel: Event-induced absorbed dose rates of the historic Carrington (dashed lines) and AD774/775 event (solid lines) based on the GLE spectrum of February 1956 (GLE05, light blue), August 1972 (GLE24, red), and October 1989 (GLE45, blue). For comparison, the absorbed dose rates of the original GLE05 (see left panel) are displayed (solid green line).}
\label{fig:8}
\end{figure*}
%
\subsubsection{Total absorbed dose rate induced by modern strong SEP events}
\label{sec:gles}
In a second step, we studied the influence of 67 out of the 72 GLE events observed throughout the ground-based instrumental era on the Venusian atmosphere. Because Venus is closer to the Sun, the GLE spectra were scaled to the Venusian orbit in order to reflect a reliable radiation hazard they might have caused within the Venusian atmosphere. We assumed that all of the investigated GLE events measured at Earth also affected the Venusian atmosphere. Because SEPs are accelerated along the heliospheric magnetic field, a good connection between the Sun and the planet is mandatory in order for the event to arrive at Venus. This might not have been the case, however. This investigation should therefore be  considered as a case study for the effect of strong SEP events on the Venusian radiation dose.

The corresponding total absorbed dose rates, the sum of both the GCR- (solid black line) and GLE-induced absorbed dose rates, are displayed as colored lines in the left panel of Fig.~\ref{fig:8}. As already pointed out by \citet{herbst2019venus}, substantial differences between the GCR- and  GLE-induced ionization profiles can be observed (their Fig. 6). As pointed out, differences of up to 11 orders of magnitudes could be observed at high altitudes. The left panel of Fig.~\ref{fig:8} shows that the differences within the CR-induced atmospheric radiation dose profiles at these altitudes are not as dramatic. Here, increases of up to six orders of magnitude can be found. Of particular interest for this study is the influence of strong SEP events on the Venusian HZ. Here, differences caused by the varying shapes of the GLE spectra can be observed. While GLE62 (4 November 2001) would have had no influence on the Venusian radiation dose, GLE05 (23 February 1956), the strongest GLE event directly measured at Earth so far, would have led to drastic increases in the absorbed radiation dose of microbial cells of two orders of magnitude within the HZ.

We note, however, that strong SEP events are often accompanied by so-called Forbush decreases (FDs), which are sudden decreases of the GCR background intensity caused by interplanetary shocks or magnetized ejecta of coronal mass ejections \citep[see, e.g.,][]{Cane-2000}. In addition to this, the GCR-induced radiation dosage would be reduced, and thus the total absorbed radiation dose a microbial cell would suffer within the HZ during strong SEP events.
\subsubsection{Total absorbed dose rate induced by historic extreme SEP events}
\begin{table*}[!t]
\begin{center}
\caption{Computed absorbed dose rates $\mean{D}$ a water-based microbial cell would suffer within a potential Venusian HZ during extreme SEP events. Results are given for the Carrington and AD774-775 event based on the GLEs measured in February 1956 (GLE05), August 1972 (GLE24), and October 1989 (GLE45).}
\label{tab:2}
\begin{tabular}{c c c c c | c c c}
\hline
\hline
               &   & \multicolumn{3}{c}{$\mean{D}_{Carrington}$ [mGy/h]}      & \multicolumn{3}{c}{$\mean{D}_{AD774/775}$ [mGy/h]}                            \\
                                   
               &  & \multicolumn{3}{c}{based on GLE spectrum of}      & \multicolumn{3}{c}{based on GLE spectrum of}                            \\
Altitude {[}km{]} & GCR & February 1956          & August 1972                      & October 1989   & February 1956          & August 1972                      & October 1989                   \\
\hline
51              &4.5$\times10^{-5}$  & 2.4$\times10^{-3}$ & 4.5$\times10^{-5}$ & 1.4$\times10^{-4}$ & 9$\times10^{-3}$ & 4.5$\times10^{-5}$ & 2$\times10^{-3}$ \\
62              &4.0$\times10^{-3}$  & 12.4 & 4$\times10^{-2}$ & 6$\times10^{-1}$ & 33.9 & 23.2 & 12.4 \\
\hline
\end{tabular}
\end{center}
\end{table*}

However, from historical records like the white-light flare observation on 1 September 1859, for instance, which was accompanied by what is known as the Carrington event, or from the records of the so-called cosmogenic radionuclides $^{10}$Be, $^{14}$C, and $^{36}$Cl, produced by interactions of CRs with the terrestrial atmosphere, we know that much stronger SEP events have occurred over the past 10,000 years. In particular, three exceptionally strong increases have been discussed in the literature: the AD774-775 event \citep[see, e.g.,][]{Miyake-etal-2012}, the AD993-994 event \citep[see, e.g.,][]{Mekhaldi-etal-2015}, and the BC660 event \citep[see][]{OHare-etal-2019}. With a production increase of more than 15 $\%$ above the GCR-induced background, the AD774-775 event is so far the strongest event ever recorded. Model studies by \citet{Kovaltsov-etal-2014} and \citet{Herbst-etal-2015} showed that the event was most likely more than one order of magnitude stronger than GLE05. 

However, no information about the spectral shape of these historical extreme events is available. Therefore, assumptions about the spectral shape and its temporal evolution have to be made. A common approach is to rescale the spectra of known recent GLE events. For example, \citet{Dartnell-etal-2015} scaled the the event spectra of GLE05 (23 February 1956), GLE24 (4 August 1972), and GLE45 (24 October 1989) to match the estimated E$>$30 MeV integral fluence of the Carrington event \citep[F$_{30}$= 2$\times 10^{10}$ protons cm$^{-2}$][]{Cliver-Dietrich-2013} and the estimated E$>200$ MeV fluence of the AD774-775 event (F$_{200}$ = 8$\times 10^{9}$ protons cm$^{-2}$) by \citet{Kovaltsov-etal-2014}. The duration of the events also has to be estimated in order to account for average differential spectra. Following the approach by \citet{Dartnell-etal-2015}, an event duration of 20 hours was assumed for both historical events. Using these spectra \citep[see, e.g., Fig. 3 in][]{Dartnell-etal-2015}, we here calculate the altitude-dependent absorbed dose rates microbial cells would suffer during such extreme solar events. 

The results of this study are shown in the right panel of Fig.~\ref{fig:8}. Here the total absorbed dose rates of the Carrington and the AD774-775 event based on the different energy spectra are displayed. With increases of up to five orders of magnitude above the GCR-induced background, these extreme events would have a tremendous impact on the absorbed dose rate values within the HZ. The AD 774-775 event rescaled ton the basis of the hard spectrum of GLE05 would have had the most substantial influence throughout the entire HZ, while the Carrington event rescaled based on the soft spectrum of GLE24 would have had almost no impact on the absorbed dose rates within the HZ.

According to \citet{Dartnell-etal-2015}, at the top of the HZ, the radiation dose of the GCR-induced absorbed dose is about 9.5$\cdot10^{-4}$ mGy/h, while the GLE05-based Carrington- and AD775 events show values of 1.6 mGy/h and 4.9 mGy/h, respectively. A comparison with the values a water-based microbial cell would suffer at the lower and the upper boundary of the potential Venusian HZ as given in Table~\ref{tab:2} shows that the radiation hazard for any putative Venusian cloud-based life is about five times higher than previously reported. This difference most likely arises because a Venusian-air-based phantom was used and the quality factors of different radiation types were therefore not accounted for.

In order to directly compare the Venusian radiation hazard on water-based life, we can compare our results to the absorbed dose rates measured at Earth. Here, the GCR-induced background leads to absorbed dose rates between 5$\times 10^{-4}$ mGy/h at the ground \citep[$\sim$ 1033 g/cm$^2$, see, e.g.,][]{banjac-etal-2018} and $0.3$ mGy/h at flight altitudes where the dose rate maximum is located \citep[$\sim$ 120 g/cm$^2$ see, e.g.,][]{Herbst-etal-2019b}. Thus, the GCR-induced Venusian absorbed dose rates at the top of the HZ, for example, are well below the magnitudes that can be found at flight altitudes at Earth. As has been pointed out by \citet{Dartnell-etal-2015}, even for the worst-case scenario of the GLE05-based AD774/775 spectrum, the induced absorbed radiation dose rates of about 33.9 mGy/h at the top of the habitable zone are in addition well below those of the impact of GLE05 on the terrestrial absorbed dose rate values \citep[41.7 mGy/h, see][]{Herbst-etal-2019b}. However, as has been pointed out by \citet{Dartnell-etal-2015}, even radiation-sensitive microorganisms, for example, \textit{Shewanella oneidensis,} can survive much higher radiation dosages than those induced during this worst-case scenario. According to \citet{Ghosal-etal-2005}, 90$\%$ of \textit{Shewanella oneidensis} cells do not survive 24 hours of being exposed to a radiation of 70 Gy. The cells can therefore survive absorbed radiation doses below 2.9 Gy/h.

Consequently, all of the so-far known extreme SEP events (AD774-775, AD 992-993, and BC660) would not have had a hazardous effect on possible microorganisms within the Venusian habitable zone. 
\section{Summary and conclusions}
Here we provided updated altitude-dependent Venusian radiation dose profiles based on the newly developed full 3D Monte Carlo simulation code AtRIS \citep{banjac-etal-2018}. We used the most recent numerical physics models (Geant4 10.5, FTFP$\_$BERT$\_$HP) as well as the primary GCR spectra of hydrogen (Z=1) to nickel (Z=28) from the CREME2009 model \citep[see, e.g.,][]{Tylka-etal-1997} with energies between 1~MeV and 10~TeV. To study the influence of strong solar events on the Venusian radiation field, we also studied the effects of 67 space-age GLE events, and two extreme historical events. 

We showed that the contribution of GCR protons and alpha particles to the induced Venusian radiation field, in particular in the HZ, are in the order of up to 75$\%$ and 20$\%$, respectively. The contribution of Z$>$2 particles is less than 10 $\%$ and can therefore be neglected. 

For the first time, we investigated the Venusian absorbed dose rates in different phantoms. Here we also took phantoms into account that are built out of water, which is the key ingredient for life as we know it from Earth. An ICRU-sphere-like liquid-water phantom, mimicking the human body, and a microbial cell of 50 $\mu$m, were also developed and implemented in AtRIS. We showed that the computations by \citet{Dartnell-etal-2015} lead to an underestimation of the absorbed dose rates that water-based organisms would suffer within the Venusian atmosphere by up to 45$\%$ for a 50$\mu$m microbial cell, and up to 60$\%$ compared for a water-based ICRU-sphere phantom. However, as discussed below, compared to the limitations of lethal radiation levels, these differences may only play a minor role.

As recently shown by \citet{herbst2019venus}, modeling the influence of strong SEP events on the Venusian atmosphere can be restricted to studying the influence of solar protons. Using the spectra of GLE events measured at Earth \citep[provided by, e.g.,] []{Raukunen-etal-2018}, we investigated the impact of the strongest SEP events measured throughout the space era (1942 to 2012) on the radiation field and thus also any putative organisms in the Venusian atmosphere. We showed that the strongest GLE event measured so far, the event of 23 February 1956, with a rather hard spectrum, would have led to a drastic increase in the absorbed dose rates of up to two orders of magnitude within the potential upper and middle habitable zone. From the cosmogenic radionuclide records of $^{10}$Be, $^{14}$C, and $^{36}$Cl, however, we know that much stronger events have occurred in the past. The dramatic production rate increases of more than 15$\%$ above the GCR-induced background around AD774-775, AD993-994, and BC660 are of particular interest in this context \citep[][ respectively]{Miyake-etal-2012, Mekhaldi-etal-2015, OHare-etal-2019}. As discussed in \citet{Nordheim-etal-2015} and \citet{herbst2019venus}, these historical events would have caused dramatic changes in the Venusian ionization rates, of about up to six orders of magnitude. A comparison to Earth-bound measurements shows that the induced Venusian radiation dose values are much lower than the induced values within the Earth's atmosphere. We showed that all known extreme SEP events would not have had a hazardous effect on potential microorganisms, and even radiation-sensitive microorganisms like \textit{Shewanella oneidensis} would have survived. This further supports the previous findings by \citet{Dartnell-etal-2015}. 

However, most recently, \citet{Notsu-etal-2019} found that old Sun-like stars can produce superflares with energies above 5$\times 10^{34}$erg (5$\times 10^{-1}$W/m$^2$) once every 2000 to 3000 years. Thus, much stronger SEP events than the imprints that have so far been found in the cosmogenic radionuclide records of $^{10}$Be, $^{14}$C, and $^{36}$Cl around AD774-5, AD992-3, and 660 BC \citep[see, e.g.,][ respectively]{Miyake-etal-2012, Mekhaldi-etal-2015, OHare-etal-2019} may have occurred on the Sun. According to \citet{Herbst-etal-2019a}, such energetic flares could lead to particle events that are more intense by than two orders of magnitude than GLE05, which would drastically increase the absorbed radiation dose values within microbial cells in the HZ of the Venusian atmosphere. Even more intense flares with energies up to 10$^{36}$ erg (10 W/m$^2$) have been detected in young G-type stars like our Sun \citep[see][]{Notsu-etal-2019}. This means that if they were produced by our Sun earlier in its history, such intense flares may have produced particle events five orders of magnitude stronger than GLE05 \citep[see][]{Herbst-etal-2019a}. These superflare events could have had drastic impacts on any potential airborne life in the Venusian clouds. 
\begin{acknowledgements}
KH and SB thank the German Research Foundation (DFG) for financial support via the project \textit{The Influence of Cosmic Rays on Exoplanetary Atmospheric Biosignatures} (Project number 282759267) and further J. L. Grenfell, B. Heber, H. Rauer, M. Scheucher, V. Schmidt, and M. Sinnhuber for a fruitful collaboration. KH and DA acknowledge the International Space Science Institute and the supported International Team 464: \textit{The  Role  Of  Solar  And  Stellar  Energetic  Particles  On (Exo)Planetary Habitability (ETERNAL, \url{http://www.issibern.ch/teams/exoeternal/})}. KH also acknowledges Team 441: \textit{High EneRgy sOlar partICle Events Analysis (HEROIC, \url{http://www.issibern.ch/teams/heroic/})}. Work by TAN was carried out at the Jet Propulsion Laboratory, California Institute of Technology, under a contract with the National Aeronautics and Space Administration.
\end{acknowledgements}
}
\bibliographystyle{aa}
\bibliography{references}

\begin{thebibliography}{62}
\expandafter\ifx\csname natexlab\endcsname\relax\def\natexlab#1{#1}\fi

\bibitem[{Allison {et~al.}(2016)Allison, Amako, Apostolakis, Arce, Asai, Aso,
  Bagli, Bagulya, Banerjee, Barrand, Beck, Bogdanov, Brandt, Brown, Burkhardt,
  Canal, Cano-Ott, Chauvie, Cho, Cirrone, Cooperman, Cortés-Giraldo, Cosmo,
  Cuttone, Depaola, Desorgher, Dong, Dotti, Elvira, Folger, Francis, Galoyan,
  Garnier, Gayer, Genser, Grichine, Guatelli, Guèye, Gumplinger, Howard,
  Hřivnáčová, Hwang, Incerti, Ivanchenko, Ivanchenko, Jones, Jun,
  Kaitaniemi, Karakatsanis, Karamitros, Kelsey, Kimura, Koi, Kurashige,
  Lechner, Lee, Longo, Maire, Mancusi, Mantero, Mendoza, Morgan, Murakami,
  Nikitina, Pandola, Paprocki, Perl, Petrović, Pia, Pokorski, Quesada, Raine,
  Reis, Ribon, Fira, Romano, Russo, Santin, Sasaki, Sawkey, Shin, Strakovsky,
  Taborda, Tanaka, Tomé, Toshito, Tran, Truscott, Urban, Uzhinsky, Verbeke,
  Verderi, Wendt, Wenzel, Wright, Wright, Yamashita, Yarba, \&
  Yoshida}]{Allison-etal-2016}
Allison, J., Amako, K., Apostolakis, J., {et~al.} 2016, Nuclear Instruments and
  Methods in Physics Research Section A: Accelerators, Spectrometers, Detectors
  and Associated Equipment, 835, 186

\bibitem[{Ball(2017)}]{Ball-2017}
Ball, P. 2017, Proceedings of the National Academy of Sciences, 114, 13327

\bibitem[{Banjac {et~al.}(2019{\natexlab{a}})Banjac, Heber, Herbst, Berger, \&
  Burmeister}]{Banjac-etal-2019b}
Banjac, S., Heber, B., Herbst, K., Berger, L., \& Burmeister, S.
  2019{\natexlab{a}}, Journal of Geophysical Research: Space Physics, n/a, n/a

\bibitem[{Banjac {et~al.}(2019{\natexlab{b}})Banjac, Herbst, \&
  Heber}]{banjac-etal-2018}
Banjac, S., Herbst, K., \& Heber, B. 2019{\natexlab{b}}, Journal of Geophysical
  Research: Space Physics, 124, 50

\bibitem[{{Bazilevskaya} {et~al.}(2008){Bazilevskaya}, {Usoskin},
  {Fl{\"u}ckiger}, {Harrison}, {Desorgher}, {B{\"u}tikofer}, {Krainev},
  {Makhmutov}, {Stozhkov}, {Svirzhevskaya}, {Svirzhevsky}, \&
  {Kovaltsov}}]{Bazilevskaya-etal-2008}
{Bazilevskaya}, G.~A., {Usoskin}, I.~G., {Fl{\"u}ckiger}, E.~O., {et~al.} 2008,
  Space Sci. Rev., 137, 149

\bibitem[{{Belov} {et~al.}(2005){Belov}, {Eroshenko}, {Mavromichalaki},
  {Plainaki}, \& {Yanke}}]{Belov-etal-2005}
{Belov}, A., {Eroshenko}, E., {Mavromichalaki}, H., {Plainaki}, C., \& {Yanke},
  V. 2005, Annales Geophysicae, 23, 2281

\bibitem[{Blamont {et~al.}(1986)Blamont, Young, Seiff, Ragent, Sagdeev, Linkin,
  Kerzhanovich, Ingersoll, Crisp, Elson, Preston, Golitsyn, \&
  Ivanov}]{Blamont-etal-1986}
Blamont, J.~E., Young, R.~E., Seiff, A., {et~al.} 1986, Science, 231, 1422

\bibitem[{Bombardieri {et~al.}(2007)Bombardieri, Michael, Duldig, \&
  Humble}]{Bombardieri-etal-2007}
Bombardieri, D.~J., Michael, K.~J., Duldig, M.~L., \& Humble, J.~E. 2007, The
  Astrophysical Journal, 665, 813

\bibitem[{Borucki {et~al.}(1982)Borucki, Levin, Whitten, \&
  Keesee}]{Borucki-etal-1982}
Borucki, W., Levin, Z., Whitten, R., \& Keesee, R. 1982, Icarus, 321, 302–321

\bibitem[{{B\"usching} {et~al.}(2005){B\"usching}, {Kopp}, {Pohl},
  {Schlickeiser}, {Perrot}, \& {Grenier}}]{Buesching-etal-2005}
{B\"usching}, I., {Kopp}, A., {Pohl}, M., {et~al.} 2005, Astrophys. J., 619,
  314

\bibitem[{Cane(2000)}]{Cane-2000}
Cane, H.~V. 2000, Space Science Reviews, 93, 55

\bibitem[{{Cliver} \& {Dietrich}(2013)}]{Cliver-Dietrich-2013}
{Cliver}, E.~W. \& {Dietrich}, W.~F. 2013, Journal of Space Weather and Space
  Climate, 3, A31

\bibitem[{Dartnell {et~al.}(2015)Dartnell, Nordheim, Patel, Mason, Coates, \&
  Jones}]{Dartnell-etal-2015}
Dartnell, L., Nordheim, T.~A., Patel, M.~R., {et~al.} 2015, Icarus, 257,
  396–405

\bibitem[{Donahue \& Hodges(1992)}]{Donahue-Hodges-1992}
Donahue, T.~M. \& Hodges, J. R.~R. 1992, Journal of Geophysical Research:
  Planets, 97, 6083

\bibitem[{Donahue {et~al.}(1982)Donahue, Hoffmann, Hodges, \&
  Watson}]{Donahue-etal-1982}
Donahue, T.~M., Hoffmann, J.~H., Hodges, R.~R., \& Watson, A.~J. 1982, Science,
  216, 630

\bibitem[{Fukuhara {et~al.}(2017)Fukuhara, Futaguchi, Hashimoto, Horinouchi,
  Imamura, Iwagaimi, Kouyama, Murakami, Nakamura, Ogohara, Sato, Sato, Suzuki,
  Taguchi, Takagi, Ueno, Watanabe, Yamada, \& Yamazaki}]{Fukuhara-etal-2017}
Fukuhara, T., Futaguchi, M., Hashimoto, G.~L., {et~al.} 2017, Nature
  Geoscience, 10

\bibitem[{Geant4\_Collaboration(2018)}]{collab2017prm}
Geant4\_Collaboration. 2018, Accessible from the GEANT4 web page [1]

\bibitem[{Ghosal {et~al.}(2005)Ghosal, Omelchenko, Gaidamakova, Matrosova,
  Vasilenko, Venkateswaran, Zhai, Kostandarithes, Brim, Makarova, Wackett,
  Fredrickson, \& Daly}]{Ghosal-etal-2005}
Ghosal, D., Omelchenko, M.~V., Gaidamakova, E.~K., {et~al.} 2005, FEMS
  Microbiology Reviews, 29, 361

\bibitem[{{Gieseler} \& {Heber}(2016)}]{Gieseler-Heber-2016}
{Gieseler}, J. \& {Heber}, B. 2016, Astron. Astrophys., 589, A32

\bibitem[{Grinspoon \& Bullock(2013)}]{Grinspoon-Bullock-2007}
Grinspoon, D.~H. \& Bullock, M.~A. 2013, Astrobiology and Venus Exploration
  (American Geophysical Union (AGU)), 191--206

\bibitem[{{Guo} {et~al.}(2019){Guo}, {Banjac}, {R\"ostel}, {Terasa}, {Herbst},
  {Heber}, \& {Wimmer-Schweingruber}}]{Guo-etal-2019}
{Guo}, J., {Banjac}, S., {R\"ostel}, L., {et~al.} 2019, J. Space Weather Space
  Clim., 9, A2

\bibitem[{Hassler {et~al.}(2014)Hassler, Zeitlin, Wimmer-Schweingruber,
  Ehresmann, Rafkin, Eigenbrode, Brinza, Weigle, B{\"o}ttcher, B{\"o}hm,
  Burmeister, Guo, K{\"o}hler, Martin, Reitz, Cucinotta, Kim, Grinspoon,
  Bullock, Posner, G{\'o}mez-Elvira, Vasavada, Grotzinger, \&
  Team}]{Hassler-etal-2014}
Hassler, D.~M., Zeitlin, C., Wimmer-Schweingruber, R.~F., {et~al.} 2014,
  Science, 343

\bibitem[{{Heber} {et~al.}(1996){Heber}, {Droege}, {Ferrando}, {Haasbroek},
  {Kunow}, {Mueller-Mellin}, {Paizis}, {Potgieter}, {Raviart}, \&
  {Wibberenz}}]{Heber-etal-1996}
{Heber}, B., {Droege}, W., {Ferrando}, P., {et~al.} 1996, Astron. Astrophys.,
  316, 538

\bibitem[{{Herbst} {et~al.}(2019{\natexlab{a}}){Herbst}, {Banjac}, \&
  {Nordheim}}]{herbst2019venus}
{Herbst}, K., {Banjac}, S., \& {Nordheim}, T.~A. 2019{\natexlab{a}}, Astron.
  Astrophys., 624, A124

\bibitem[{{Herbst} {et~al.}(2019{\natexlab{b}}){Herbst}, {Grenfell},
  {Sinnhuber}, {Rauer}, {Heber}, {Banjac}, {Scheucher}, {Schmidt}, {Gebauer},
  {Lehmann}, \& {Schreier}}]{Herbst-etal-2019b}
{Herbst}, K., {Grenfell}, J., {Sinnhuber}, M., {et~al.} 2019{\natexlab{b}},
  Astron. Astrophys., 631

\bibitem[{{Herbst} {et~al.}(2015){Herbst}, {Heber}, {Beer}, \&
  {Tylka}}]{Herbst-etal-2015}
{Herbst}, K., {Heber}, B., {Beer}, J., \& {Tylka}, A.~J. 2015, The 34th
  International Cosmic Ray Conference, The Hague, Proceedings of Science (PoS),
  537

\bibitem[{Herbst {et~al.}(2013)Herbst, Kopp, \& Heber}]{Herbst-etal-2013}
Herbst, K., Kopp, A., \& Heber, B. 2013, Annales Geophysicae, 31, 1637

\bibitem[{{Herbst} {et~al.}(2019{\natexlab{c}}){Herbst}, Papaioannou, Banjac,
  \& {Heber}}]{Herbst-etal-2019a}
{Herbst}, K., Papaioannou, A., Banjac, S., \& {Heber}, B. 2019{\natexlab{c}},
  Astron. Astrophys., 621

\bibitem[{{Hillas}(2005)}]{Hillas-2005}
{Hillas}, A.~M. 2005, Phys. G: Nucl. Part. Phys., 31, R95

\bibitem[{ICRP(2007)}]{ICRP-2007}
ICRP. 2007, ICRP Publication 103. Ann. ICRP, 37, 2

\bibitem[{{ICRP 60}(1991)}]{protection1991icrp}
{ICRP 60}. 1991, Ann. ICRP, 21

\bibitem[{Keating {et~al.}(1985)Keating, Bertaux, Bougher, Dickinson, Cravens,
  Nagy, Hedin, Krasnopolsky, Nicholson, Paxton, \& von
  Zahn}]{Keating-etal-1985}
Keating, G., Bertaux, J., Bougher, S., {et~al.} 1985, Advances in Space
  Research, 5, 117

\bibitem[{{Kliore} {et~al.}(1992){Kliore}, {Keating}, \&
  {Moroz}}]{Kliore-etal-1985}
{Kliore}, A.~J., {Keating}, G.~M., \& {Moroz}, V.~I. 1992, \planss, 40, 573

\bibitem[{{Kovaltsov} \& {Usoskin}(2014)}]{Kovaltsov-etal-2014}
{Kovaltsov}, G.~A. \& {Usoskin}, I.~G. 2014, \solphys, 289, 211

\bibitem[{Limaye {et~al.}(2018)Limaye, Mogul, Smith, Ansari, Słowik, \&
  Vaishampayan}]{Limaye-etal-2018}
Limaye, S.~S., Mogul, R., Smith, D.~J., {et~al.} 2018, Astrobiology, 18, 1181

\bibitem[{{Matthi{\"a}} {et~al.}(2009){Matthi{\"a}}, {Heber}, {Reitz}, {Meier},
  {Sihver}, {Berger}, \& {Herbst}}]{Matthiae-etal-2009}
{Matthi{\"a}}, D., {Heber}, B., {Reitz}, G., {et~al.} 2009, J. Geophys. Res.:
  Space Phys., 114, A08104

\bibitem[{McNair(1981)}]{icru-1980}
McNair, A. 1981, Journal of Labelled Compounds and Radiopharmaceuticals, 18,
  1398

\bibitem[{{Mekhaldi} {et~al.}(2015){Mekhaldi}, {Muscheler}, {Adolphi},
  {Aldahan}, {Beer}, {McConnell}, {Possnert}, {Sigl}, {Svensson}, {Synal},
  {Welten}, \& {Woodruff}}]{Mekhaldi-etal-2015}
{Mekhaldi}, F., {Muscheler}, R., {Adolphi}, F., {et~al.} 2015, Nature
  Communications, 6, 8611

\bibitem[{{Miroshnichenko}(2018)}]{Miroshnichenko-2018}
{Miroshnichenko}, L.~I. 2018, Journal of Space Weather and Space Climate, 8,
  A52

\bibitem[{Miyake {et~al.}(2012)Miyake, Nagaya, Masuda, \&
  Nakamura}]{Miyake-etal-2012}
Miyake, F., Nagaya, K., Masuda, K., \& Nakamura, T. 2012, Nature, 486, 240

\bibitem[{Morales-Olivares \&
  Caballero-Lopez(2010)}]{Morales-Olivares-Caballero-Lopez-2010}
Morales-Olivares, O. \& Caballero-Lopez, R. 2010, Advances in Space Research,
  46, 1313

\bibitem[{Nilsson-Almqvist \& Stenlund(1987)}]{Nilsson-Almqvist-Stenlund-1987}
Nilsson-Almqvist, B. \& Stenlund, E. 1987, Computer Physics Communications, 43,
  387

\bibitem[{Nordheim {et~al.}(2015)Nordheim, Dartnell, Desorgher, Coates, \&
  Jones}]{Nordheim-etal-2015}
Nordheim, T.~A., Dartnell, L., Desorgher, L., Coates, A., \& Jones, G. 2015,
  Icarus, 245, 80–86

\bibitem[{Notsu {et~al.}(2019)Notsu, Maehara, Honda, Hawley, Davenport,
  Namekata, Notsu, Ikuta, Nogami, \& Shibata}]{Notsu-etal-2019}
Notsu, Y., Maehara, H., Honda, S., {et~al.} 2019, The Astrophysical Journal,
  876, 58

\bibitem[{O{\textquoteright}Hare {et~al.}(2019)O{\textquoteright}Hare,
  Mekhaldi, Adolphi, Raisbeck, Aldahan, Anderberg, Beer, Christl, Fahrni,
  Synal, Park, Possnert, Southon, Bard, , \& Muscheler}]{OHare-etal-2019}
O{\textquoteright}Hare, P., Mekhaldi, F., Adolphi, F., {et~al.} 2019,
  Proceedings of the National Academy of Sciences, 116, 5961

\bibitem[{Petoussi-Henss {et~al.}(2010)Petoussi-Henss, Bolch, Eckerman, Endo,
  Hertel, Hunt, Pelliccioni, Schlattl, \& Zankl}]{Petoussi-Henss-etal-2010}
Petoussi-Henss, N., Bolch, W., Eckerman, K., {et~al.} 2010, Annals of the ICRP,
  40, 1

\bibitem[{{Pfotzer}(1936)}]{Pfotzer-1936}
{Pfotzer}, G. 1936, Z. Phys, 102, 41

\bibitem[{Phillips \& Russell(1987)}]{Phillips-etal-1987}
Phillips, J.~L. \& Russell, C.~T. 1987, J. Geophys. Res.: Space Phys., 92, 2253

\bibitem[{Picone {et~al.}(2002)Picone, Hedin, Drob, \&
  Aikin}]{picone2002nrlmsise}
Picone, J., Hedin, A., Drob, D.~P., \& Aikin, A. 2002, Journal of Geophysical
  Research: Space Physics, 107

\bibitem[{Plainaki {et~al.}(2007)Plainaki, Belov, Eroshenko, Mavromichalaki, \&
  Yanke}]{Plainaki-etal-2007}
Plainaki, C., Belov, A., Eroshenko, E., Mavromichalaki, H., \& Yanke, V. 2007,
  Journal of Geophysical Research: Space Physics, 112

\bibitem[{Plainaki {et~al.}(2016)Plainaki, Paschalis, Grassi, Mavromichalaki,
  \& Andriopoulou}]{Plainaki-etal-2016}
Plainaki, C., Paschalis, P., Grassi, D., Mavromichalaki, H., \& Andriopoulou,
  M. 2016, Annales Geophysicae, 34, 595

\bibitem[{{Raukunen} {et~al.}(2018){Raukunen}, {Vainio}, J., {Dietrich},
  {Jiggens}, {Heynderickx}, {Dierckxsens}, {Crosby}, {Ganse}, \&
  {Siipola}}]{Raukunen-etal-2018}
{Raukunen}, O., {Vainio}, R., J., T.~A., {et~al.} 2018, J. Space Weather Space
  Clim., 8, A04

\bibitem[{{Reames}(1999)}]{Reames-1999}
{Reames}, D.~V. 1999, Space Sci. Rev., 90, 413

\bibitem[{Russell {et~al.}(2012)Russell, Elphic, \& Slavin}]{Russell-etal-1980}
Russell, C.~T., Elphic, R.~C., \& Slavin, J.~A. 2012, J. Geophys. Res., 85,
  8319–8332

\bibitem[{{Schmelz} {et~al.}(2012){Schmelz}, {Reames}, {von Steiger}, \&
  {Basu}}]{Schmelz-etal-2012}
{Schmelz}, J.~T., {Reames}, D.~V., {von Steiger}, R., \& {Basu}, S. 2012,
  Astrophys. J., 755, 33

\bibitem[{Schulze-Makuch {et~al.}(2004)Schulze-Makuch, Grinspoon, Abbas, Irwin,
  \& Bullock}]{Schulze-Makuch-etal-2004}
Schulze-Makuch, D., Grinspoon, D.~H., Abbas, O., Irwin, L.~N., \& Bullock,
  M.~A. 2004, Astrobiology, 4, 11

\bibitem[{Seiff {et~al.}(1985)Seiff, Schofield, Kliore, Taylor, Limaye,
  Revercomb, Sromovsky, Kerzhanovich, Moroz, \& Marov}]{Seiff-etal-1985}
Seiff, A., Schofield, J., Kliore, A., {et~al.} 1985, Advances in Space
  Research, 5, 3

\bibitem[{Simon~Wedlund {et~al.}(2011)Simon~Wedlund, Gronoff, Lilensten,
  M\'enager, \& Barth\'elemy}]{Wedlund-etal-2011}
Simon~Wedlund, C., Gronoff, G., Lilensten, J., M\'enager, H., \& Barth\'elemy,
  M. 2011, Annales Geophysicae, 29, 187

\bibitem[{{Taylor} {et~al.}(2018){Taylor}, {Svedhem}, \&
  {Head}}]{Taylor-etal-2018}
{Taylor}, F.~W., {Svedhem}, H., \& {Head}, J.~W. 2018, \ssr, 214, 35

\bibitem[{Tylka {et~al.}(1997)Tylka, Adams, Boberg, Brownstein, Dietrich,
  Flueckiger, Petersen, Shea, Smart, \& Smith}]{Tylka-etal-1997}
Tylka, A.~J., Adams, J.~H., Boberg, P.~R., {et~al.} 1997, IEEE Transactions on
  Nuclear Science, 44, 2150

\bibitem[{Valentin {et~al.}(2007)}]{valentin2007icrp}
Valentin, J. {et~al.} 2007, The 2007 recommendations of the international
  commission on radiological protection (Elsevier Oxford)

\bibitem[{Way {et~al.}(2016)Way, Del~Genio, Kiang, Sohl, Grinspoon, Aleinov,
  Kelley, \& Clune}]{Way-etal-2016}
Way, M.~J., Del~Genio, A.~D., Kiang, N.~Y., {et~al.} 2016, Geophysical Research
  Letters, 43, 8376

\end{thebibliography}
\end{document}